\documentclass[journal]{IEEEtran}
\pdfoutput=1
\usepackage[utf8]{inputenc}
\usepackage[T1]{fontenc}

\usepackage{color}
\usepackage{amsmath,amssymb}
\usepackage{bm}
\usepackage{multirow}

\usepackage{mathtools}

\usepackage{tikz}
\usepackage{pgfplots}
\usepackage{pgfplotstable}
\usepackage{caption}
\usepackage{subcaption}
\usepackage{float}
\usepackage{lipsum}
\usepackage{multicol}
\usepackage{graphicx}
\usepackage[ruled,vlined, linesnumbered]{algorithm2e}

\usetikzlibrary{math}
\tikzmath
{
  function symlog(\x,\a){
    \yLarge = ((\x>\a) - (\x<-\a)) * (ln(max(abs(\x/\a),1)) + 1);
    \ySmall = (\x >= -\a) * (\x <= \a) * \x / \a ;
    return \yLarge + \ySmall ;
  };
  function symexp(\y,\a){
    \xLarge = ((\y>1) - (\y<-1)) * \a * exp(abs(\y) - 1) ;
    \xSmall = (\y>=-1) * (\y<=1) * \a * \y ;
    return \xLarge + \xSmall ;
  };
}

\usepgfplotslibrary{groupplots}
\pgfplotsset{compat=1.12,
SmallBarPlot/.style={
    font=\footnotesize,
    ybar,
    width=\linewidth,
    ymin=0,
    xtick=data,
    xticklabel style={text width=1.5cm, rotate=90, align=center}
},
BlueBars/.style={
    fill=blue!20, bar width=0.25
},
RedBars/.style={
    fill=red!20, bar width=0.25
},
select coords between index/.style 2 args={
    x filter/.code={
        \ifnum\coordindex<#1\fi
        \ifnum\coordindex>#2\fi
    }
}
}

\usetikzlibrary{automata, positioning, arrows}
\tikzset{
>=stealth,
node distance=3cm, % specifies the minimum distance between two nodes. Change if necessary.
every state/.style={thick, fill=gray!10}, % sets the properties for each ’state’ node
initial text=$ $, % sets the text that appears on the start arrow
}

\usepackage{algorithmic}
\usepackage{listings}
\usepackage{url}
\usepackage{enumitem}
\usepackage{booktabs}
\ifCLASSINFOpdf
\else
\fi

\usepackage[textsize=tiny, textwidth=1.35cm]{todonotes}

\usepackage[normalem]{ulem}
\newcommand{\change}[2]{\textcolor{red}{\sout{#1}}\textcolor{blue}{#2}}
\renewcommand{\change}[2]{\textcolor{blue}{#2}}

\newcommand{\stkout}[1]{\ifmmode\text{\sout{\ensuremath{#1}}}\else\sout{#1}\fi}

\usepackage[hidelinks, breaklinks]{hyperref}

\begin{document}
% \setlength{\marginparwidth}{1.35cm}

% Extra page to fix the issue with package \texttt{todonotes} and double-click-to-go-to-source feature.
% \clearpage
% \setcounter{page}{1}

%
% paper title

\title{Reinforcement Learning based Proactive Control for Transmission Grid Resilience to Wildfire}
\author{S.~U.~Kadir,~\IEEEmembership{Student~Member,~IEEE,}
        S.~Majumder,~\IEEEmembership{Member,~IEEE,} A. Chhokra, ~\IEEEmembership{Member,~IEEE,} A. Dubey,~\IEEEmembership{Senior~Member,~IEEE,} H. Neema,  A. Laszka, and~A.~Srivastava,~\IEEEmembership{Senior~Member,~IEEE}% <-this % stops a space
\thanks{S. U. Kadir and A. Laszka are with the University of Houston, Houston, TX.}
\thanks{S. Majumder and A. Srivastava are with the
%School of Electrical Engineering and Computer Science,
Washington State University, Pullman, WA (e-mail: anurag.k.srivastava@wsu.edu)%
%99163 USA e-mail: (anurag.k.srivastava@wsu.edu)
.}% <-this % stops a space
\thanks{A. Chhokra, A. Dubey and H. Neema are with the Vanderbilt University, Nashville, TN.}% <-this % stops a space
\thanks{Authors acknowledge that this work was in part supported by National Science Foundation (NSF) awards 1840192, 1840083 and 1840052.}}

\maketitle

% As a general rule, do not put math, special symbols or citations
% in the abstract or keywords.

% For peer review papers, you can put extra information on the cover
% page as needed:
% \ifCLASSOPTIONpeerreview
% \begin{center} \bfseries EDICS Category: 3-BBND \end{center}
% \fi
%
% For peerreview papers, this IEEEtran command inserts a page break and
% creates the second title. It will be ignored for other modes.
\IEEEpeerreviewmaketitle

% \section{Introduction}
% % The very first letter is a 2 line initial drop letter followed
% % by the rest of the first word in caps.
% % 
% % form to use if the first word consists of a single letter:
% % \IEEEPARstart{A}{demo} file is ....
% % 
% % form to use if you need the single drop letter followed by
% % normal text (unknown if ever used by the IEEE):
% % \IEEEPARstart{A}{}demo file is ....
% % 
% % Some journals put the first two words in caps:
% % \IEEEPARstart{T}{his demo} file is ....
% % 
% % Here we have the typical use of a "T" for an initial drop letter
% % and "HIS" in caps to complete the first word.
% \IEEEPARstart{T}{his} demo file is intended to serve as a ``starter file''
% for IEEE journal papers produced under \LaTeX\ using
% IEEEtran.cls version 1.8b and later.
% % You must have at least 2 lines in the paragraph with the drop letter
% % (should never be an issue)
% I wish you the best of success.

\begin{abstract}

Power grid operation subject to an extreme event requires decision-making by human operators under stressful condition with high cognitive load. Decision support under adverse dynamic events, specially if forecasted, can be supplemented by intelligent proactive control. Power system operation during wildfires require resiliency-driven proactive control for load shedding, line switching and resource allocation considering the dynamics of the wildfire and failure propagation. However, possible number of line- and load-switching in a large system during an event make traditional prediction-driven and stochastic approaches computationally intractable, leading operators to often use greedy algorithms.
%The complexity of proactive control is increased because it requires
%The problem complexity is further increased with the necessity of the 
%integration of the wildfire progression model. 
We model and solve the proactive control problem
%To enable the training of such an intelligent agent, we pose the control problem
as a Markov decision process and introduce an integrated testbed for spatio-temporal wildfire propagation and proactive power-system operation. We transform the enormous wildfire-propagation observation space and utilize it as part of a heuristic for proactive de-energization of transmission assets. We integrate this heuristic with a reinforcement-learning based proactive policy for controlling the generating assets. Our approach allows this controller to provide setpoints for a part of the generation fleet, while a myopic operator can determine the setpoints for the remaining set, which results in a symbiotic action. We evaluate our approach utilizing the IEEE 24-node system mapped on a hypothetical terrain. Our results show that the proposed approach can help the operator to reduce load loss  during an extreme event, reduce power flow through lines that are to be de-energized, and reduce the likelihood of infeasible power-flow solutions, which would indicate violation of short-term thermal limits of transmission lines.

\end{abstract}

% Note that keywords are not normally used for peerreview papers.

% \input{1_introduction}
% \section{Nomenclature}
% \textcolor{red}{Move symbols here}.
\vspace{-6pt}
\section{Introduction} 

\IEEEPARstart{T}{he} frequency of extreme weather events (also classified as high-impact low-frequency, or HILF, events), such as storms, floods, and wildfires, as well as the associated impact on the power grid have soared in recent years \cite{USbillion-dollar}. Changing climate raises the potential for frequent wildfire events. Furthermore, increased ambient temperature due to approaching wildfires, or heatwaves in general, can result in sags, expanding the potential of faults for power transmission lines \cite{CHOOBINEH201520}. 
Additionally, high winds can blow nearby vegetation into transmission lines, which can create sparks and snap transmission lines with the potential of originating secondary wildfire hazards \cite{Western.2021}. To prevent these secondary sources, several utilities in the USA typically restrict
%have adopted the strategy to restrict the
power flow through some of their assets during emergency events  (e.g., public safety power shut-off, or PSPS, events in the state of California, USA~\cite{abatzoglou2020population}). Compared to other extreme weather events, the slow progression of wildfires~\cite{7105972} across the large geographical span of power systems provides an opportunity for proactive control. In particular, slow-moving nature of these events provides  operators with sufficient time to actively control line flows before the preemptive de-energization of power-system assets. Here, we define proactive control as any pre-event or during-event action to minimize the expected impact of an evolving extreme event.

From an operational point of view, if a heavily loaded line suffers an outage due to sags or if transmission lines exceed their short-term overload capacity, the necessary power flow redistribution may lead to a cascaded outage (as observed in the 1977 New York black-out \cite{Time2015NewYork}). Since the cost of power-system restoration can be enormous, it is crucial to reroute transmission flow with necessary de-energization proactively, while ensuring that consumers experience the minimum inconvenience of power shut off. 
While the use of traditional proactive control strategies for power-system resiliency is abundant \cite{8894433}, most of them revolve around controlling the operation of the distribution network. 
However, the bulk of power still flows through the transmission network, and while periodic vegetation management around the transmission infrastructure may help, wildfire threats exist nonetheless. Hence, the proactive control of transmission assets has to be
emphasized. Eventually, this will help operators to avoid  power-system operational reliability issues, while indirectly reducing the chances of secondary wildfires and event impact propagation.

% Compared to traditional power system operational control approaches, literature on the resilient transmission network operation is limited. In this domain, 
Traditional power transmission system operation is governed by power system economics and the $N-1$ operational reliability criterion. However, if the emergency warning and caution (EWAC) direction is received from various stakeholders, operators typically aim to predict outage risks to take necessary control measures to minimize loss of loads, and  systems will move to resiliency mode. Techniques, such as Markovian model-based proactive sequential re-dispatch of generators \cite{wang2016resilience}, proactive splitting the transmission grid into islands \cite{biswas2020proactive,6344599}, transmission system reconfiguration \cite{8053185}, the coordinated control of multiple microgrids connected via transmission system \cite{7857787, 7127029}, defensive islanding formation \cite{7434044}, optimal power shut-offs \cite{rhodes2020balancing}, transmission line derating \cite{CHOOBINEH201520,10.1007/978-3-662-48768-6_139}, are common approaches for resilient transmission network operation and control. 
Abundant monitoring data from SCADA and PMUs can be  leveraged {in} %determining the 
outage {forecasting}, {which can be  utilized} for proactive resource allocation and  {deployment of necessary} measures to serve  {consumers} during emergency conditions. In this regard, statistical models of outage duration 
% \cite{doi:10.1061/(ASCE)1076-0342(2005)11:4(258), 6949604, nateghi2014forecasting} 
\cite{doi:10.1061/(ASCE)1076-0342(2005)11:4(258), 6949604} can be  applied by the operators. 
However, most of these techniques employ classical optimization-based approaches and are combinatorial in nature, making the control task computationally challenging and resource-intensive to handle in real-time. Furthermore, recent events have shown the unsuitability of traditional forecast models in an effective determination of outage risk {(e.g.,  majority of the wind power generators were outaged due to lack of winterization in the 2021 Texas power outage event, and this was not captured in the generation forecast} \cite{Texas.2021}). 
Likewise, traditional model-predictive control techniques are also quite resource-intensive for real time.
% \Aron{this sentence is disconnected from the rest}
% Additionally, the use of predicted wildfires propagation path into an early warning signal for the transmission system outage has been considered in \cite{8620983}. 
%Once the emergency conditions are lifted, the system returns to normal economics-driven operation, limiting the scope of our research. \ad{I dont get the point of this last sentence. How is this connected with rest of the paragraph. if the paragraph is about related research then say what is missing.} \sm{This is to stress that emergency control will be activated after EWAC is received. And hence, the scope of research is limited.}
%\Aron{unclear to me too, can we just omit?}
%
%
% THERE WAS A PARAGRAPH BREAK HERE
%
\change{In a deregulated environment, during the normal operating mode, or during the emergency condition, the control actions obtained would require proper coordination among various agents, such as generating entities, Independent System Operators (ISOs) / Regional Transmission Operators (RTOs), Distribution System Operators (DSOs), aggregators, and load-serving entities. Available abundant monitoring data from SCADA and PMUs can be appropriately leveraged for determining the outage risk following EWAC vis-\'a-vis for proactive resource allocation and measures to serve the customers during emergency condition. In this regard, statistical models of outage duration 
% \cite{doi:10.1061/(ASCE)1076-0342(2005)11:4(258), 6949604, nateghi2014forecasting}
(e.g. \cite{nateghi2014forecasting}) can be suitably utilized by the operators. However, the combinatorial nature of the solution space, especially with optimal line-switching and generation control \cite{1372745,4162603,ahmadi2014distribution}, 
% \cite{1372745,4162603,ahmadi2014distribution}, 
make this task computationally challenging to handle in real-time. \change{Alternative}{Additionally,} traditional model predictive control techniques are also quite resources intensive.}{}
%
%
% THERE WAS A PARAGRAPH BREAK HERE
%
\change{While one can also utilize the existing wildfire}{} 
% I COMMENTED THIS OUT SINCE THE PARAGRAPH HAS BEEN REMOVED -Aron
%\ad{this whole paragraph is disconnected in the sense that we need a statement first describing the gap in related research and then state the approach is to use predictive models and decision theory. However, the prediction models themselves are not available readily.. This will also help your contribution summary... } 
\change{propagation dataset (refer to \cite{neurips20} for one such sample wildfire dataset) along with historical power system monitoring and control data for determining suitable control actions using machine learning (ML) approaches,}{}

Alternatively, machine learning (ML) approaches (e.g., \cite{nateghi2014forecasting}) can be suitably integrated by operators into proactive decision-making. ML models that were trained \emph{apriori} can be leveraged to significantly reduce  computation time during an event, offloading most of the decision-making onto the trained controller, which results in fast and efficient decision-making \cite{7951075}. Wildfire propagation data (see ref. \cite{neurips20} for one such sample dataset) along with historical power-system monitoring and control data can be leveraged to this end. The controller can complement the action taken by the power system operator, resulting in an efficient operation. This also alleviates the difficulties of traditional proactive optimization and rule-based approaches while minimizing the possibility of human errors due to operating in a stressful environment. In this regard, reinforcement learning (RL) based power-system decision support has gained significant traction in recent years \cite{chen2021reinforcement}. This, along with recent advances in RL-based control \cite{KHARGONEKAR20181}, indicate the plausibility of successful deployment of ML-based techniques for proactive control in the advent of a disaster.

% \textcolor{red}{https://ayanmukhopadhyay.github.io/publications/neurips20 -- lets add this here}
 
% In this work, we have tried to address the following queries:

% \begin{itemize}
%     \item \textbf{Can we use artificial intelligence to assist the power system operator in avoiding the potential cascading during a natural disaster like wildfire?} In the simulation, we have seen that the power system operator can avoid the cascading with the help of the AI agent. 
%     \item \textbf{What is the extent of the assistance if AI agent can help the operator?} We have found that it works better if the AI agent works together with the power system operator; instead AI agent controls the whole power generation. 
%     \item \textbf{What is the possible difficulties to build such an intelligent agent? And how did we resolve it?} We faced several difficulties with building such an intelligent agent. The main obstacle was designing and training the agent for large and multidimensional observation and action spaces. To solve the issue, we followed several techniques to reduce the observation and action spaces. We propose a hybrid model which combines the heuristic rules with the AI agent.     
% \end{itemize}

%This research aims to build an external automated agent that will be able to help the power system operator during a natural disaster like wildfire to protect the power system component and keep the load loss minimum. The core contribution of this paper is as follows:
In this paper, we propose a novel approach to aid  power-system operators in effective proactive control of available resources during wildfires and develop an associated simulation testbed for training and validation. The core contributions of this paper are as follows:

\begin{itemize}
\item Modeled the spatio-temporal dynamics of a wildfire and the operation of a transmission systems as a Markov decision process, which captures the impact of the wildfire on the power-system assets. 
Our model allows the decision-support agent to control generation setpoints and de-energize lines working in tandem with existing power system operation and associated controller.% can take inputs from existing controllers with flexibility.
\item Formulated the optimal proactive control problem for minimizing load shedding, the de-energization of  power-system assets proactively, and the likelihood of infeasible power-flow solutions, considering the time horizon of the wildfire event. Proactive de-energization of line avoids outage of live line and impact propagation.
\item  %we propose a}{
Proposed a novel approach, which is an ensemble of a compact representation for the agent's observation of the wildfire state, a heuristic algorithm for the proactive de-energization of power lines, and a deep reinforcement learning based approach trained using Deep Deterministic
Policy Gradient (DDPG) for the proactive control of power generation. %\change{}{The proposed approach aims to solve the optimal proactive control problem efficiently.}
\item Developed an integrated testbed combining a wildfire and power-system simulator (available as open-source in the future). Testbed was used to train and evaluate developed agent on the IEEE transmission system test system mapped onto a topographical map. %\footnote{We will make our testbed implementation available as open-source software.}
%\Aron{add a sentence saying that }
We demonstrate that our proactive approach can significantly reduce system impact compared to reactive and myopic control policies.
\end{itemize}

\section{Model Description and Problem Formulation}
\label{sec:model}
% \begin{figure}
%     \centering
%     \includegraphics[width=\columnwidth]{figures/pics/blockdiagram.pdf}
%     \caption{System Block Diagram }
%     \label{fig:blockdiagram}
% \end{figure}

\begin{figure*}
    \centering
    \includegraphics[width=2\columnwidth]{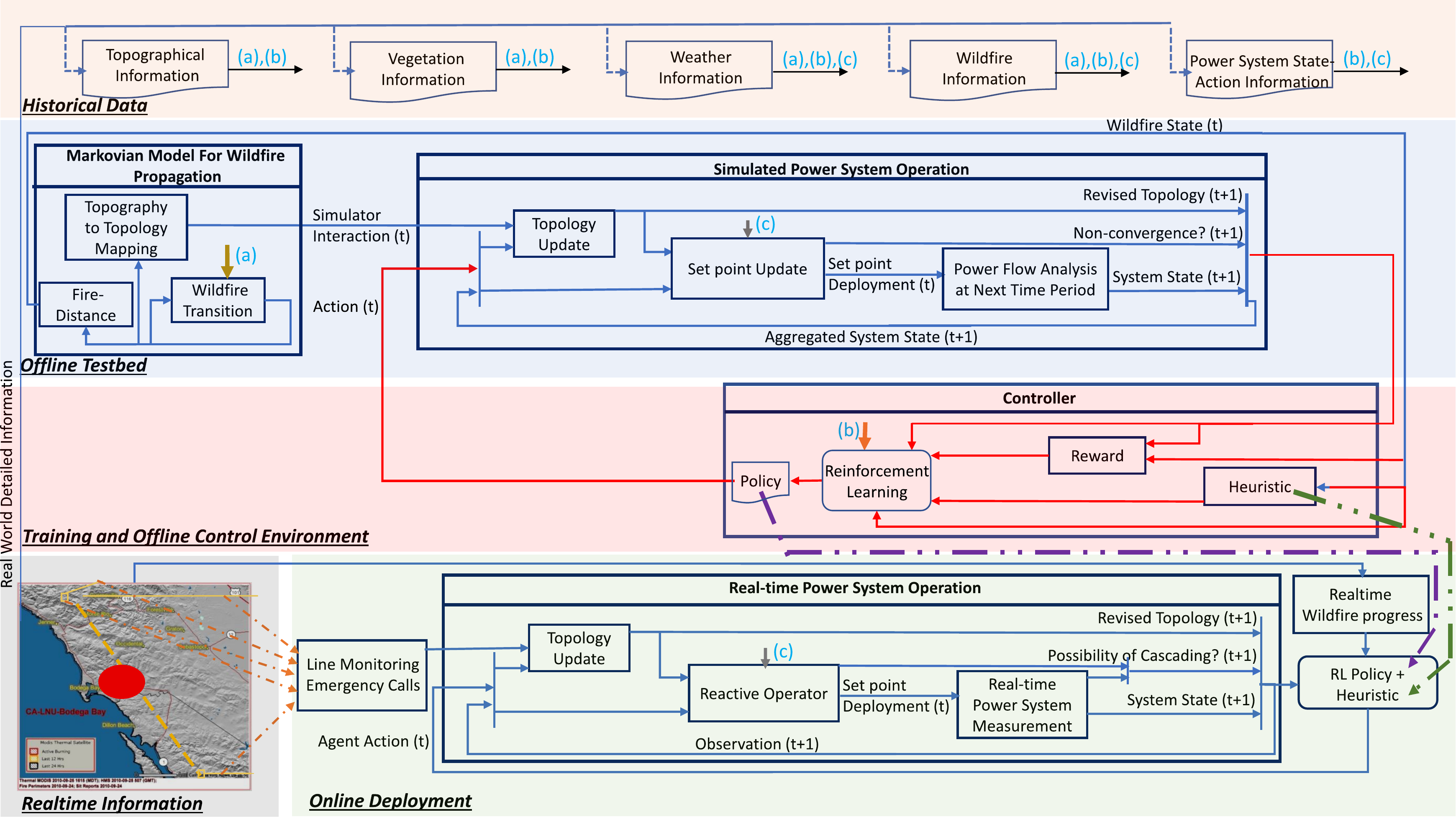}
    \caption{An integrated wildfire propagation and reactive power system operation --- testbed, training, and deployment.}
    \label{fig:blockdiagram}
\end{figure*}

% \Aron{need a better opening sentence (or sentences)}
Fig.~\ref{fig:blockdiagram} shows the high-level overview of the integrated power-system operation supplemented by an external controller to avoid catastrophic damage caused by the wildfire events modeled as a Markovian decision process (MDP). The integrated method comprises four major components: (i) historical data, (ii) offline integrated testbed, (iii) training and offline control environment, and (iv) online deployment of the controller in the natural environment. Here, accumulated historical data can be suitably leveraged to develop the wildfire transition model and also by the operator both during the training and real-time operation. All these historical datasets can also be leveraged for the development of a crude controller for further training. This section provides a detailed treatment of wildfire propagation and power-system operation modeling and formulates the control problem as MDP. A detailed discussion of the controller and associated training will be provided in the following
section.

\subsection{Wildfire Model}
\subsubsection{Wildfire Propagation Dynamics in Topographical Space}

\label{sec:wildfire}
We utilize a stochastic model as given in ref. \cite{Bertsimas2017AProblems} for wildfire propagation dynamics. We divided the geographical region into multiple cells and defined it by $X$ as a grid cell set. The temporal horizon is also divided into uniform-length contiguous steps identified by the variable $k$. The state $s^{f}_{x,k}$ of fire in each cell $x \in X$ at the beginning of $k^{th}$ interval (i.e., at the $k^{th}$ time step) is captured by a Boolean variable $d^x_k$, representing the status of fire, and an integer variable $h^x_k$ ($\geq 0$), representing available wildfire fuel within the cell (determining the duration of fire).

The status of fire $d^x_k$ in cell $x$ can be either one  of two states:  non-ignited $d_k^x = 0$ and  ignited $d_k^x = 1$. Once a cell is ignited, it consumes fuel at a constant rate of C$^x$ until it exhausts (burns) all   fuel in the cell. {Precise  fuel dynamics are given by \eqref{eq:fire3}. Once all the fuel is burnt, the cell returns to the non-ignited state.}

\begin{equation}
    h^{x}_{k} = 
    \begin{cases}
        h^{x}_{k-1} & \text{if} \quad \neg d^{x}_{k-1} \lor h^{x}_{k-1} \leq 0 \\ 
        h^{x}_{k-1} - \text{C}^{x}     & \text{otherwise.} 
    \end{cases}
    \label{eq:fire3}
\end{equation}

The evolution of $d^{x}_k$ is stochastic and  driven by the transition probability $\rho_{k}^{x}$. 
%
% 
% \Salah{again feel like blackbox. It seems to me we can simply remove this line.}
% \Aron{as I have suggested before, remove this figure because it shows something trivial and takes up about 9 lines of space}
% Figure~\ref{fig:bx} illustrates the probabilistic transition model for $d^{x}_{k}$ where $\rho_{k}^{x}$ is the probability of the cell $x$ being ignited at  $k^{th}$ time step  is given by \eqref{eq:fire4}. 
% the geographical and environmental factors are implicitly captured by the wildfire propagation transitional probability and is known a priori (could be calculated from historical information)
Specifically, the probability of  cell $x$ being ignited at $k^{th}$ time step (i.e., $d^x_k = 1$) is given by~\eqref{eq:fire4}.
\begin{equation}
    \rho^{x}_{k} = \begin{cases}
    0                                                       & \text{if} \quad \neg d^{x}_{k-1} \land |\mathcal{H}^x_k| = 0 \\
    1 - \prod_{y \in \mathcal{H}^{x}_{k}}\: (1 - \text{P}^{y}_{x,k}) &  \text{if} \quad \neg d^{x}_{k-1} \land |\mathcal{H}^x_k| > 0 \\
    1                                                       & \text{otherwise.} 
        \end{cases}
    \label{eq:fire4}
\end{equation}

Here, $\text{P}^{y}_{x,k}$ denotes the probability of fire spreading from cell $y$ to $x$. For a given cell $x$, $\mathcal{H}^{x}_{k}$ is the set of neighboring cells that can contribute to the spreading of fire to cell $x$ in $k^{th}$ time step, that is, $\mathcal{H}^{x}_{k}$ is the set of neighboring cells $y$ for which $d_k^y = 1$. %\change{}{(only the neighbouring cells that are burning, given by $d^x_k$, can contribute to wildfire progress; and also if the concerning cell is already burnt, or has no wildfire fuel, given by $h^{x}_{k}$, fire cannot encroach into it)}.
% as each cell, $y \in \mathcal{H}^{x}_{k}$ was burning in the previous time step i.e. $d_{k-1}^y  \:\:\land\:\: h_{k-1}^{y} > 0$ (\texttt{IGNITED}). 
Note that  geographical and environmental factors are implicitly captured by the cell-to-cell wildfire spread probability $\text{P}^{y}_{x,k}$.  {Real world} topographical information can be stored as historical data and utilized to determine or update these probabilities. However, the calculation of these probabilities is beyond the scope of this paper, and we assume them to be given. 
% \Salah{feel like blackbox, it seems to me that it could be better having a little hints or any reference.}

As shown in \eqref{eq:fire2}, the state of the whole topographical grid $s^{f}_k$ can be {obtained} by composing the state of every cell.

\begin{equation}
\begin{gathered}
    s^{f}_{k} = \left\langle (s_{1,k}) , \ldots , (s_{M,k}) \right\rangle \\
    \mathcal{L}(s^{f}_{k}) = \left\langle d^1_k , \ldots , d^M_k \right\rangle
\end{gathered}
\label{eq:fire2}
\end{equation}
where $(1, ..., M)$ are $M$ cells in the grid. 

Note  that the selection of a larger spatiotemporal grid would speed up the simulation of the fire-propagation model at the expense of a significant reduction in model accuracy~\cite{boychuk2009stochastic}. Therefore, it is assumed that the fire propagation model operates on a finer timescale, and the result of the simulation is appropriately down-sampled for determining the outcome of wildfire-power system interactions. The selection of optimal grid size is beyond the scope of this paper.

% \begin{figure}
%     \centering
%     \begin{tikzpicture}
%         \node[state, initial] (q1) {False};
%         \node[state, right of=q1] (q2) {True};
%         \draw   
%                 (q1) edge[loop above] node{$1 - \rho^x$} (q1)
%                 (q1) edge[above] node{$\rho^x$} (q2);
% \end{tikzpicture}
%     \caption{Transition model for $B^x$} 
%     \label{fig:bx}
% \end{figure}

\subsubsection{Topography to Topology Mapping}
\label{sec:impact}

Given the wildfire dynamics in Section \ref{sec:wildfire}, here we model the impact of propagating wildfire on the power system operation. The power system environment is geotagged with  cell information, and the state of one or more cells---in terms of $\mathcal{L}(s^x_k)$---correspond to a given power system asset {or equipment (transmission lines or substations)}. Topologically, the power system can be represented as a collection of nodes, $N$, representing generation and transmission substations, connected through a set of transmission lines, $T$. At a given time step, for each node $i \in N$ and branch $t \in T$, given cell-level fire propagation status obtained earlier, we assign binary variables $z_{i,k}^{f}$ and $z_{t,k}^{f}$ to indicate the operational status of substations and transmission lines respectively ($= 1$ denotes availability, and $= 0$ denotes assets encroached by fire) and is captured using \eqref{eq:impact}.

\begin{equation}
        z_{(\cdot),k}^{f} = 
        \begin{cases}
        0 & \text{if} \quad  \exists x \in G_{(\cdot)} \:\:\text{s.t.}\:\: \mathcal{L}(s^x_k) \ne \text{non-ignited}  \\
        1 & \text{otherwise}
    \end{cases}
    \label{eq:impact}
\end{equation}
where set $G_{(\cdot)}$ symbolizes the cells corresponding to a given power system asset. Although the wildfire propagation model operates on a finer time scale compared to power system operation, we reuse the variable $k$ to determine power system operational time steps without the loss of generality. 
% As indicated in Fig.~\ref{fig:blockdiagram} \Salah{there is no indication in figure 1}, $\lbrace z_{i,k}^f~\forall i \in N ,~z_{t,k}^f~\forall t \in T \rbrace$ constitutes the simulator interaction provided from wildfire propagation block to simulated power system operation block. 
In the real-world deployment, this input mimics the emergency call from the generating and transmission substations, or thermal overload condition of the transmission lines \cite{ltmABB} based on temperature measurement, indicating wildfire encroachment.

\subsection{Power System Operation Model}
\label{sec:power}

Contrary to an existing energy management system (EMS) providing decision-support to the operator in solving traditional multi-period scenario-based deterministic or stochastic operational optimization problem for the decision making, as shown in Fig.~\ref{fig:blockdiagram}, we treat our system operator as a  myopic entity, actively seeking the recommendation of the controller in the decision-making in the wake of the disaster, while monitoring the state of the system along with the system-wide emergency condition. Since the applicability of the proposed controller is limited to the wildfire time horizon, and the operator is expected to return to economic mode following expiration emergency conditions, the operating horizon of the controller is finite. Operator actions can be divided into two~stages:

\subsubsection{Topology Revision} We consider that the action taken by the operator is also discrete in time, and is based on the observation made at the beginning of $k^{th}$ interval, or at $k^{th}$ time step. Here, the operator receives the emergency responses from the wildfire simulation block, given by $z_{i,k}^{f}$ and $z_{t,k}^{f}$ (see \eqref{eq:impact}), and substation and transmission line shut-off actions from the controller, given as $z_{i,k}^{e}$ and $z_{t,k}^{e}$ respectively, and first deploys them faithfully\footnote{The operator may choose to not strictly follow the decision support provided by the controller. However, such an operator needs to be appropriately modeled to ensure controller robustness.}. Thus, transmission asset switching actions are not automatic, rather, those actions will be supervised by the operator. Consequently,\change{given the state of the power system at $k^{th}$ time step,}{} the revised operational status of a substation and transmission line for the ${k+1}^{th}$ time step becomes $z^{o}_{i,k+1} = z_{i,k}^{e}z_{i,k}^{f}z^{o}_{i,k}$ and $z^{o}_{t,k+1} = z_{t,k}^{e}z_{t,k}^{f}z^{o}_{t,k}$ respectively (where $= 1$ denotes availability, and $= 0$ denotes shut-off condition of the transmission assets).

\subsubsection{Reactive Operator Actions} Given the updated system topology ($z^{o}_{i,k+1}$ and $z^{o}_{t,k+1}$), existing system state ($s^p_k$), and partial generator control setpoints from the controller at $k^{th}$ time step, the operator also utilizes the demand forecast to derive the complete list of setpoints of the generators and load shedding schedule while aiming to minimize the value of the lost load at $k+1^{th}$ time step. As indicated earlier, contrary to traditional economics-driven power system operation, we consider that the operator's objective should be to minimize the value of the shredded load at $k+1^{th}$ time step\footnote{The operator may also use a multi-period greedy algorithm to derive and deploy set points for next time step.}, and such an objective will remain in effect until the emergency condition is lifted. The use of such an objective is often notable in the post-blackout power distribution system restoration problem~\cite{8421054}. The consequent problem to be solved is given as:
\begin{equation}
    \min_{\Phi_{k+1}} \sum_{i \in N} w^{c\_l}_i \Delta P^{c\_l}_{i,k+1} + w^{nc\_l}_i \Delta P^{nc\_l}_{i,k+1}  + \epsilon \left( P^{g}_{i,k+1} -  P^{g}_{i,k} \right)^2
    \label{eq:objective}
\end{equation}
where $\Phi_{k+1}$ consists of the set of decision variables of the operator, that includes power generation outputs $P^{g}_{i,k+1}$ (without any loss of generality, all the generators at a given bus are aggregated), operational state of generators group $z_{i,k+1}^g$, critical, non-critical loads shedded ($\Delta P^{c\_l}_{i,k+1}$, $\Delta P^{nc\_l}_{i,k+1}$), nodal angles $\theta_{i,k+1}$, of node $i \in N$ along with line flows, $P^{flow}_{t,k+1}$, through branch $t \in T$. Also, $w^{c\_l}_i$, $w^{nc\_l}_i$ $(> 0)$ are the values of critical and non-critical loads respectively. The criticality of the loads (hospitals, fire stations, police stations, etc. are generally treated as critical loads) imposes the condition of $w^{c\_l}_i \geq w^{nc\_l}_i > 0$ ($\forall i \in N$). These values can be different for different nodes. The nullity of the decision variable space $\Phi_{k+1}$ indicates non-existence of the solution, implying, possible overloading and outage of transmission resources along with secondary wildfire origin, and is captured through variable $q_{k+1}$. Overall state of the power system is captured by $s^{p}_{k+1}$ = \{$\Phi_{k+1} \cup q_{k+1} \cup z^{o}_{t,k+1} \cup z^{o}_{i,k+1}$\}.

Sole utilization of the value of load loss expression as an objective, $\sum_{i \in N} w^{c\_l}_i \Delta P^{c\_l}_{i,k+1} + w^{nc\_l}_i \Delta P^{nc\_l}_{i,k+1}$, may lead to the existence of multiple feasible solutions with undesirable ramping of the generators. Addition of an expression $\sum_{i \in N} \left( P^{g}_{i,k+1} -  P^{g}_{i,k} \right)^2$ in the objective with minuscule positive bias of $\epsilon$ inhibits such possibility, with little to no impact on the original objective.  Furthermore, $\min{w^{nc\_l}_i} \gg \epsilon > 0$. Determination of these weights are beyond the scope of this paper, and is assumed to be known \emph{apriori}. 

The discussed objective function must be subject to the following power system operation and safety constraints:

\paragraph{Generation Constraints}
Finite generation capacity (lower and upper limits of generators given by P$^{min}_i$, P$^{max}_i$, respectively) and ramping capabilities (given by $\text{R}^{max}_{i}$) limit the operability of the generators. \change{These parameters are positive real numbers.}{} Since the network topology has already been revised, we use $z^{o}_{i,k+1}$, and $z^{o}_{t,k+1}$ to determine nodal and line availability, respectively. The controller provides the list of `to be controlled' generators through $v^{e}_{i,k}$ ($= 1$ symbolizes to be controlled, $=0$ otherwise), and their incremental setpoint adjustment, $\Delta P^{e}_{i,k}$. The operator, in return, checks the feasibility in terms of upper limits of the received set-point based on the current operating state of the generators $z^{g}_{i,k}$ using \eqref{eq:in:su} ($= 1$ symbolizes operational, $=0$ otherwise), failing which the operator reports nullity of the decision space $q_{k+1}$.
\begin{equation}
        0 \leq v^{e}_{i,k} \left( P^{g}_{i, k} + \Delta P^{e}_{i,k} \right) \leq v^{e}_{i,k} z^{o}_{i,k+1} z^{g}_{i,k} \text{P}^{max}_i \label{eq:in:su} 
\end{equation}

If the lower limit is violated (as indicated by $z^e_{i,k} = 0$ in \eqref{eq:in:su3}), the limits remain unchanged; otherwise, minimum and maximum operating limits of the generators ($\text{P}^{min,*}_i, \text{P}^{max,*}_i$) are updated using \eqref{eq:in:su1}-\eqref{eq:in:su2}.
\begin{gather}
        z^{e}_{i,k} = \begin{cases}
            0 & \text{if}~~ v^{e}_{i,k} \left( P^{g}_{i, k} + \Delta P^{e}_{i,k} \right)  \leq v^{e}_{i,k} z^{o}_{i,k+1} z^{g}_{i,k} \text{P}^{min}_i \\ 
            1 & \text{otherwise}
                             \end{cases}   \label{eq:in:su3}\\
        \text{P}^{min,*}_i = z^{e}_{i,k} \left( P^{g}_{i, k} + \Delta P^{e}_{i,k} \right) +  (1 - z^{e}_{i,k}) \text{P}^{min}_i \label{eq:in:su1}\\
        \text{P}^{max,*}_i = z^{e}_{i,k} \left( P^{g}_{i, k} + \Delta P^{e}_{i,k} \right) +  (1 - z^{e}_{i,k}) \text{P}^{max}_i  \label{eq:in:su2}
\end{gather}

In a continuously changing environment, generators can suffer from forced outages. Here, we assume that the generators are equipped with load rejection capabilities \cite{4111968}, and when generators face forced outages, their ramping rate and operating limit can be allowed to contravene (see \eqref{eq:Gen_lim}), subject to the outaged generators will not be brought online without a thorough safety check. For this paper, these generators will remain outaged indefinitely (see \eqref{eq:zo}). To model such a condition, we seek the help of a large positive real constant, $\Gamma^0$, as shown in \eqref{eq:ramp_lim}. These conditions, along with revised generating and ramping capability, are given in the following equations. Here, $\Delta k$ represents the power system operating interval.
\begin{gather}
        0 \leq z^{g}_{i,k+1} \leq z^{g}_{i,k} \label{eq:zo} \\
        z^{g}_{i,k+1} {\text{P}^{min,*}_{i}} \leq P^{g}_{i,k} \leq z^{g}_{i,k+1} {\text{P}^{max,*}_{i}} \label{eq:Gen_lim} \\
        -\Gamma^{0} \left(1-z^{g}_{i,k+1}\right) -\Delta k R^{max}_{i} \leq P^{g}_{i,k+1} - P^{g}_{i,k}~~~~~~~~~~~~~~~~~~~~~~~~~ \nonumber \\
        ~~~~~~~~~~~~~~~~~~~~~~~~~~\leq \Delta k R^{max}_i + \Gamma^{0} \left(1-z^{g}_{i,k+1}\right) \label{eq:ramp_lim}
\end{gather}

\paragraph{Load Demand Constraints}
The necessity of the deployment of control action for the generators at $k^{th}$ time step to ensure load-generation balance at $k+1^{th}$ time step requires prediction of load demand. Available historical data, as shown in Fig.~\ref{fig:blockdiagram}, can facilitate such computation. However, the development of load prediction models is also beyond the scope of this paper and is assumed to be given.

Limited availability of generation during the prevailing contingencies necessitates demand curtailment. Suppose, P$^{l}_{i,k+1}$ is the operator predicted load demand, then, following removal of associated substation, updated load demand will be $z^{o}_{i,k+1}\text{P}^{l}_{i,k+1}$. Additionally, $\alpha_i$ is the parameter representing the critical load fraction (positive real number) served. Availability of a large number of switchable loads within the distribution network connected at the transmission substation enables treating the sheddable loads as a continuous variable \cite{9019635}. Consequently, critical and non-critical sheddable loads ($\Delta P^{c\_l}_{i,k+1}, \Delta P^{nc\_l}_{i,k+1}$) is bounded as follows:
\begin{gather}
    0 \leq \Delta P^{c\_l}_{i,k+1} \leq \alpha_i z^{o}_{i,k+1} \text{P}^l_{i, k+1} \label{eq:Crit_lim:SM}\\
    0 \leq \Delta P^{nc\_l}_{i,k+1} \leq  (1 - \alpha_i) z^{o}_{i,k+1} \text{P}^l_{i,k+1} \label{eq:NCrit_lim:SM}
\end{gather}

\paragraph{Load Flow Constraints}
Since the typical operating voltage within the power system remains close to 1.00 \textit{pu}, and the difference in the voltage angle of the adjacent buses is tiny, we consider a DC power flow model \cite{stevenson1975element}. An associated mathematical expression is given in \eqref{eq:Power_Flow_01:SM}. Here, B$_t$ is the element corresponding to the $t^{th}$ branch in the imaginary part of the nodal admittance matrix. Equation \eqref{eq:Node_Bal:SM} represents the nodal flow balance equation. Also, the set $T^{i} \subseteq T$ consists of all the branches that are connected to node $i$. Here, $\theta^{min}$ and $\theta^{max}$ are upper and lower bound of nodal angle, respectively. The power flow constraint is described in \eqref{eq:line_therm:SM}.
\begin{gather}
    P^{g}_{i,k+1}  - \text{P}^l_{i,k+1}z^{o}_{i,k+1} + \Delta P^{c\_l}_{i,k+1} + \Delta P^{nc\_l}_{i,k+1} \nonumber \\ 
    ~~~~~~~~~~~~~~~~~~~~~~~~~~~~~~~~- \sum_{t \in T^i} P^{flow}_{t,k+1} = 0 \label{eq:Node_Bal:SM} \\
    P^{flow}_{t,k+1} -  z^{o}_{t,k+1} \text{B}_{t} \left( \theta_{i,k+1} - \theta_{j,k+1} \right)  = 0 \label{eq:Power_Flow_01:SM} \\
    \theta^{min} \leq \theta_{i,k+1} \leq \theta^{max} \label{eq:angle_lim:SM} \\
    - z^{o}_{t,k+1} \text{P}^{maxflow}_{t}  \leq P^{flow}_{t,k+1} \leq z^{o}_{t,k+1} {\text{P}^{maxflow}_{t}} \label{eq:line_therm:SM}
\end{gather}

\subsubsection{Power Flow Analysis} At each time step, the deenergization actions, along with calculated setpoints determined earlier, are deployed, and the operator waits for the duration of $\Delta k$. It is notable that the delay in the measurement of the state and deployment of the control action is minuscule enough to account for. It is also assumed that the availability of sufficient load-following reserve can help mitigate intra-period slower fluctuation \cite{doi:https://doi.org/10.1002/0470020598.ch5}. Since the forecasted system operating conditions (load demands\footnote{Renewable generation uncertainties can also be treated similarly.}) may not materialize, we need to solve an actual set of power flow equations\footnote{AC power flow equations can be invoked here.} to determine the system state. As shown in the online deployment part of Fig.~\ref{fig:blockdiagram}, the correct system state can be directly obtained from the real environment. This revised state is fed to the controller to determine requisite operator action for the next time step.

\subsection{Reward Function and MDP  Formulation}
\label{sec:Formulation}

% \Aron{agent is part of the power system operator in practice}
In our proposed model, the power system operator is  myopic. As a result, the controller needs to consider future trajectory and provide an appropriate control signal to the operator, facilitating prevention from running into {reliability} related issues while maximizing the value of load served. Given the probabilistic nature of wildfire propagation and power system loads, we formulate the control problem solved by an agent as a Markov decision process (MDP) problem. An MDP is a tuple $\mathcal{D} = \langle S, A, \mathcal{P}, \mathcal{R}\rangle$, where $S$ is a finite state space, $A$ is a finite action space, $\mathcal{P}$ is the transition probability function and $\mathcal{R} $ is the reward function. The agent chooses an action from the possible action space to lead the system from one state to another. For the given problem, these elements are defined as follows:

\subsubsection*{\textbf{States}} 
At any time step $k$, the state $s^\mathcal{E}_{k}$ of $\mathcal{D}$ is defined by a tuple $s^\mathcal{E}_{k} = \langle s^{f}_k, s^{p}_k\rangle$, including the states of wildfire and power system respectively. The complete state space $S$ is defined in~\eqref{eq:ob}. 
\begin{equation}
    S \subseteq \underbrace{\{0,1\}^{M} \times \mathbb{R}^{M}}_{S^{f} (\text{wildfire})} \times \underbrace{\{0,1\}^{ \vert T \vert + \vert N \vert +1 } \times \mathbb{R}^{\vert T \vert + 3\vert N \vert}}_{S^{p} (\text{power system}) }
    \label{eq:ob}
\end{equation} 
As discussed in Section \ref{sec:wildfire}, $M$ is the number of cells for depicting the topographical map for wildfire propagation, $\vert N \vert$ is the number of substations, and $\vert T \vert$ is the numbers of transmission lines. We assume that the state of the environment is entirely observable to the controller (i.e., to the agent).

% \textbf{States:} At any time step $k$, the complete state space, $S$, of the system comprises of wildfire and power system states, and is a subset of $\{0,1\}^{M + \vert T \vert + 2\vert N \vert +1 } \times \mathbb{R}^{M + \vert T \vert + 3\vert N \vert}$. As discussed in Section \ref{sec:wildfire}, $M$ is the number of cells for depicting the topographical map for wildfire propagation, $\vert N \vert$ is the number of substations and $\vert T \vert$ is the numbers of transmission lines. We assume that the state of the environment is completely observable.

\subsubsection*{\textbf{Actions}} 
The agent can provide emergency shut off commands to the substations and transmission lines ($z^{e}_{i,k}$, $z^{e}_{t,k}$, respectively), and change power injections for the generators $v^{e}_i$, by an amount $\Delta P^{e}_{i}$. The action space $A$ is defined as $A \subseteq \{0,1\}^{2\vert N \vert + \vert T \vert} \times \mathbb{R}^{\vert N \vert}$.

\subsubsection*{\textbf{Transitions}} 
An MDP evolves as a result of the set of actions taken. The transition probability function, $\mathcal{P}(s_{k+1}|s_k,a_k)$ indicates that the action $a_k$ at time step $k$ in state $s_k$ will lead to the next state $s_{k+1}$. Here, the stochasticity arises from the wildfire propagation model and variation in the load demand (along with generation uncertainty with renewable energy resources) within the power system.

\subsubsection*{\textbf{Reward Function}}
The reward function, $r_k = \mathcal{R}(s_k, a_k, s_{k+1})$, indicates that the reward is received for taking action $a_k$ in state $s_k$ to reach next state $s_{k+1}$. 
% Since the goal of the agent is to minimize load loss along with prevention of possible cascading, we define a reward function that imposes penalties if the power system operator incurs a load loss, enforced removal of energized assets through the selected action, fail to provide preventive action to avoid an emergency condition of power system assets, and non-existence of feasible solution space following a selected action, which indicates cascading outage. 
The reward function is formally defined by \eqref{eq:total}, where $c_1$, $c_2$, $c_3$ and $c_4$ are positive, constant weights of each types of penalty. Determining the optimal set of these weights is beyond the scope of this paper but is expected to have a major impact on the {training speed of the agent}. 
% \change{In this paper, we use the term \textit{penalty} to describe the positive value of the negative reward.} 
% {Here, \textit{penalty} represents negative reward.}

% \begin{gather}
%         \mathcal{R}(s_k, a_k, s_{k+1}) =
%         -1 \times total\ penalty \label{eq:total} \\
%         total\ penalty \:=\:  c_1r_1 + c_2r_2 +  c_3r_3 + c_4r_4\label{eq:total_penalty}  \\
%         r_1 = q_k \sum_{i \in N} \bigg( (1 - z^{o}_{i,k})\text{P}^l_{i,k} + \Delta P^{c\_l}_{i,k} + \Delta P^{nc\_l}_{i,k} \bigg) \label{eq:loadloss} \\      
%         r_2 = q_k \bigg( \sum_{i \in N} \psi^{e}_{i,k} \vert P^{g}_{i,k} \vert + \sum_{t \in T}  \psi^{e}_{t,k} \vert P^{flow}_{t,k} \vert \bigg) \label{eq:active}\\
%       r_3 = q_k \bigg( \sum_{i \in N}  \psi^{f}_{i,k+1} +  \sum_{t \in T}  \psi^{f}_{t,k+1} \bigg) \sum_{i \in N}\text{P}^l_{i,k} \label{eq:wildfire}\\ 
%       r_4 = (\vert N \cup T \vert) (1 - q_k) \sum_{i \in N}\text{P}^l_{i,k} \label{eq:constraints}
% \end{gather}

\begin{gather}
        \mathcal{R}(s_k, a_k, s_{k+1}) =
        -1 \times \left( c_1r_1 + c_2r_2 +  c_3r_3 + c_4r_4 \right)\label{eq:total}  \\
        r_1 = q_k \sum_{i \in N} \bigg( (1 - z^{o}_{i,k})\text{P}^l_{i,k} + \Delta P^{c\_l}_{i,k} + \Delta P^{nc\_l}_{i,k} \bigg) \label{eq:loadloss} \\      
        r_2 = q_k \bigg( \sum_{i \in N} \psi^{e}_{i,k} \vert P^{g}_{i,k} \vert + \sum_{t \in T}  \psi^{e}_{t,k} \vert P^{flow}_{t,k} \vert \bigg) \label{eq:active}\\
      r_3 = q_k \bigg( \sum_{i \in N}  \psi^{f}_{i,k+1} +  \sum_{t \in T}  \psi^{f}_{t,k+1} \bigg) \sum_{i \in N}\text{P}^l_{i,k} \label{eq:wildfire}\\ 
      r_4 = (\vert N \cup T \vert) (1 - q_k) \sum_{i \in N}\text{P}^l_{i,k} \label{eq:constraints}
\end{gather}

We defined the \textit{load loss} penalty $r_1$ in \eqref{eq:loadloss}, where the expression $(1 - z^{o}_{i,k})\text{P}^l_{i,k}$ signifies the shedded load demand following isolation of a substation, and $\Delta P^{c\_l}_{i,k} + \Delta P^{nc\_l}_{i,k}$ denotes aggregated critical and non-critical load curtailment of substation $i$. \change{The convergence status $q_k$ is true for all types of penalty, but false for the non-convergence penalty.}{}

The \textit{Proactive Isolation of grid Assets with expected Wildfire (PIAW)} penalty $r_2$ is proportional to the amount of power flowing through the transmission lines $ P^{flow}_{t,k}$ or the substations $P^{g}_{i,k}$ during the isolation of live assets as defined in \eqref{eq:active}, where $\psi^{e}_{j,k} = (z^{o}_{j, k} \oplus z^{e}_{j,k})$ denotes the change in the operational status of the equipment $j \in \{N \cup T\}$ as a result of controller action. Here, $\psi^{e}_{j,k}$ evaluates to true if asset $j$ status changes from $k^{th}$ time step to $k+1^{th}$ time step due to the {agent} action at time step $k$. This penalty stems from the fact that isolation of heavily loaded assets has comparably higher chance of creating power system {reliability} related problems.
% \footnote{The expression, $(z^{o}_{j, k} \oplus z^{e}_{j,k})$ evaluates to logical \emph{true} if an equipment is deenergized. %However, this is not possible by definition.
% } 

The \textit{Asset Damage and Isolation due to encroached Wildfire (ADIW)} penalty $r_3$ is defined in \eqref{eq:wildfire}, where the agent is penalized in proportion to aggregated load demand $\sum_{i \in N}\text{P}^l_{i,k+1}$. Here, $\psi^{f}_{j,k} = (z^{o}_{j, k} \oplus z^{f}_{j,k+1})$ denotes the change in the operational status of the equipment $j \in \{N \cup T\}$ as a result of safety related action following encroached wildfire. {This penalty is to prohibit safety related actions after assets are encroached by wildfire.}

The \textit{non-convergence} penalty $r_4$, defined in \eqref{eq:constraints}, is equal to the aggregated network-wide load $\sum_{i \in N}\text{P}^l_{i,k+1}$ times the number of assets $\vert N \cup T \vert$ in the system. It is notable that while non-convergence may not result into cascading outage, we use this term as a negative feedback to the controller to inhibit thermal overloading. Detailed model of slow cascading will be considered in our future research. \change{The non-convergence penalty $\vert r_4 \vert$ is the highest followed by the ADIW penalty $\vert r_3 \vert$ as the decision process must learn to satisfy all system safety constraints and the high value of ADIW penalty ensures that the decision process provides pro-active measures and should not allow wild-fire to come in contact with an energized equipment.}{}

\subsubsection*{\textbf{Policy}}
A policy $\mu(s)$ is a function that specifies the action to be taken in each state $s$. At every time step $k$, the agent selects an action, $a_k = \mu(s_k)$, based on the deployed policy. This experiment aims to find an optimal policy that can maximize the cumulative sum of expected rewards in each episode. The expected rewards in the model are defined in \eqref{eq:policy}, where $n$ is the finite number of steps in each episode. We consider the power system operator has no inherent temporal preference as the episodes' length is short (system operates in resiliency mode only for a relatively short period of time).

\begin{equation}
    \mu^* = {\arg\max}_{\mu}\: \mathbb{E}\left[ \sum_{k=0}^{n} \mathcal{R}(s_k, a_k, s_{k+1})\right] \label{eq:policy}
\end{equation}

\change{ So, we aim to maximize the undiscounted rewards over a finite time horizon that the duration of natural disaster, specifically when we expect the agent's help with proactive decisions.}{}

\section{Deep Reinforcement Learning Based Proactive Control}
\label{DRL_based_control}

We apply deep reinforcement learning to train the agent. 
We consider a standard reinforcement learning approach, where the agent interacts with its environment in discrete time steps~$k$, as illustrated in Fig. \ref{fig:blockdiagram}.
The agent learns by iteratively updating its policy $\mu^*$, defined in \eqref{eq:train_policy}, where $\gamma \in (0, 1)$ is  a temporal discount factor for  infinite-horizon future rewards:
\begin{equation}
    \mu^* = {\arg\max}_{\mu}\: \mathbb{E}\left[ \sum_{k=0}^{\infty} \gamma^k \cdot \mathcal{R}(s_k, a_k, s_{k+1})\right] \label{eq:train_policy}
\end{equation}
{Our control problem, modeled as an MDP, is fully observable, so the state $S$ and observation spaces $O$ are equivalent.} However, due to the enormous observation $O$ and action spaces~$A$ defined in Section \ref{sec:Formulation}, training {our} agent can be difficult and time consuming. To improve the agent's learning pace, we propose switching to a low-dimensional, ``compressed'' observation space $\hat{O}$. %\change{ by removing its redundancy}. 
Since the removal of an important feature to compress the observation space may also impair the agent's performance, we need to choose the compressed space $\hat{O}$ carefully. %move the action space far away from the optima.} 
{To reduce the action space,} we propose a hybrid approach, where % \change{}{to reduce the action space}. The 
control actions are partially chosen by a heuristic, inspired by \cite{rhodes2020balancing}, and by a neural network.

% \Aron{talk about fundamental challenges: observation and action spaces are so large, that even representing a policy is infeasible (hence, we will need neural networks as approximators); in fact, even with an NN approximator, it would be challenging to work with policies due to large parameter spaces (hence, we need to reduce observation space); however, observation space reduction may decrease the agent's performance as the process becomes partially observable (hence, we need a good reduction); etc.}. In the following subsections, we describe the transformations to reduce the dimensions associated with the observation space and a \Aron{I like the term `hybrid agent', we should use this in the abstract and introduction (when we talk about our contributions)}hybrid agent to tackle large action space.

\subsection{Reducing the Observation Space}

% At any given time, the observation for the agent consists of full state of the environment. The corresponding observation space, $O$, as defined in Section~\ref{sec:Formulation} can be re-arranged to separate the observation sub-space of wildfire, $S^f$ and power system $S^p$ as shown in \eqref{eq:ob}

% \begin{equation}
%     O \subseteq \underbrace{\{0,1\}^{M} \times \mathbb{R}^{M}}_{S^{f} (\text{wildfire})} \times \underbrace{\{0,1\}^{ \vert T \vert + \vert N \vert +1 } \times \mathbb{R}^{\vert T \vert + 3\vert N \vert}}_{S^{p} (\text{power system}) }
%     \label{eq:ob}
% \end{equation}
% \Aron{since we don't have much space, let's not repeat this (almost identical formulations appear in the preceding section; if you need this specific arrangement, then just include this in the preceding section)}
% where $M$ is the number of cells and ($N$, $T$) are the sets of nodes and branches in power system. 

{A}s described in Section \ref{sec:Formulation}, the %dimension of the 
complete observation space is $O \subseteq \{0,1\}^{M + \vert T \vert + \vert N \vert +1 } \times \mathbb{R}^{M + \vert T \vert + 3\vert N \vert}$. 
In practice, the number of cells $M$ in the {wildfire propagation model} is far greater than the number of substations $N$ and transmission lines $T$ (i.e., $M \gg \vert N \cup T\vert$). 
So, we can reduce the size of the observation space $O$ by reducing only the dimension of wildfire observation $S^f \subseteq \{0, 1\}^M \times \mathbb{R}^M$ %through the  {proposed} fire distance metric, 
without significantly affecting the agent's accuracy. 

To this end, we propose to replace cell states  in the observation with a ``fire-distance metric'' for each asset:
%In this transformation, we calculate  
for each power system {asset} in $N \cup T$ (substation or transmission line), calculate and observe the geographical distance {from the nearest ignited cell}. 
This transformation reduces the wildfire observation space from $S^f$ to $\hat{S}^f \subseteq \mathbb{R}^{|T|+|N|}$.
The reduced observation space $\hat{O}$ of the agent is given in \eqref{eq:rob}:
\begin{equation}
    \hat{O} \subseteq \underbrace{ \mathbb{R}^{\vert T \vert + \vert N \vert}}_{{\hat{S}^f} (\text{wildfire})} \times \underbrace{\{0,1\}^{ \vert T \vert + \vert N \vert +1 } \times \mathbb{R}^{\vert T \vert + 3\vert N \vert}}_{S^{p} (\text{power system}) }
    \label{eq:rob}
\end{equation}

% is defined in Algorithm~\ref{alg:transform}. 
% \Aron{this algorithm could be described in two-three sentences (which would probably be easier to understand than the pseudo-code)}
% The transformation produces a set of non-negative numbers, $\hat{s}^f$ where $j^{th}$  element, $\hat{s}^j \in \hat{s}^f$, represents the minimum distance of fire from $j^{th}$ equipment in power system. 

% \SetKwInput{KwInput}{Input}
% \SetKwInput{KwOutput}{Output}

% \begin{algorithm}[]
% \SetAlgoLined
% \KwIn{$s^{f}_k$, $s^{f}_{k-1}$, $M$, $N$, $T$, $\hat{s}^{f}_{k-1}$}
% \KwOut{$\hat{s}^{f}_k$}
% $ NewCells \leftarrow \varnothing$\;
% \ForEach{$x \in M$}{
%     \If{$\mathcal{L}(s^{x}_{k-1}) = \text{non-ignited}$ }{
%         \If{$\mathcal{L}(s^{x}_{k}) =  \text{ignited}$}{
%             $NewCells \leftarrow$ \{ $x \:\cup NewCells$ \}\;
%         }
%     }
% }
% \ForEach{$j \in N \cup T$}{
%     $d \leftarrow \hat{s}^j_{k-1}$\;
%     \ForEach{$x \in NewCells$}{
%         $d \leftarrow \min\{d, \texttt{distanceFromElement}(j,x)\}$\;
%     }
%     $\hat{s}^{j}_{k} \leftarrow d $\;
% }
 
%  \caption{Observation Space Transformation ($\mathcal{I}$)}
%  \label{alg:transform}
% \end{algorithm}

\subsection{Reducing the Action Space}
\label{reducing_actions}

The impact of large action space, $\{0,1\}^{2\vert N \vert + \vert T \vert} \times \mathbb{R}^{\vert N \vert}$, can be significantly reduced through direct utilization of fire-distance metric in the decision making through a simple heuristic. This heuristic can control the energization status of the component of the power system. At the same time, the neural network-trained RL setting can determine the generator's setpoints. Such a hybrid approach can alleviate the monolithic agent's bottleneck while emphasizing learning a relatively good power generation control policy. However, the action space for the RL agent ($\hat{A} \subseteq \{0,1\}^{\vert N \vert} \times \mathbb{R}^{\vert N \vert}$) is still large and we use two strategies to further scale down the action space.

% The agent can turn on or off power system equipment along with injecting power in certain nodes. The resulting actions space, $A \subseteq \{0,1\}^{2\vert N \vert + \vert T \vert} \times \mathbb{R}^{\vert N \vert}$ is too big for a \Aron{monolithic? interpretation of term may not be clear until next sentence}monolithic agent to learn a relatively good policy about. We employ a hybrid approach in which turning off a power system equipment is based on a heuristic rule and the actions related to changing  the power injections at generating substations is taken by a neural network trained in RL setting such that the new action space reduces to, $\hat{A} \subseteq \{0,1\}^{\vert N \vert} \times \mathbb{R}^{\vert N \vert}$, 
% % where $\{0,1\}^{\vert N \vert}$ represents the possibilities of selecting nodes and $\mathbb{R}^{\vert N \vert}$ possible injection values. 
% However, this action space is still large and we use two strategies to further scale down the action space.

\begin{itemize}
    \item The first strategy, referred to as \textit{full-control}, forces the agent to control all generator outputs{. Here, the} external control input $v^{e}_{i,k}$ is defined per \eqref{eq:subspace1}, and the resulting action space reduces to $\hat{A}_{1} \subseteq  \mathbb{R}^{\vert N^{gen} \vert}$, where $N^{gen} \subseteq N$ is the set of nodes with generation capabilities.
    \begin{equation}
         v^{e}_{i,k} = 
    \begin{cases}
        1 & \text{if} \quad i \in N^{gen} \\ 
        0 & \text{otherwise.} 
    \end{cases}
    \label{eq:subspace1}
    \end{equation}
    \item The second strategy, referred to as \textit{partial-control},  allows the agent to control exactly $g$ number of generators. {Due to the nature of the power transmission network, {mostly the} generators that are in the  vicinity of the wildfire need to be actively controlled. Therefore,} we can select $g$ generators by dividing the {set of} generating nodes $N^{gen}$ into $g$ {number of} {subsets}, and select one generator  from each subset. %{It is evident that the generators grouping is primarily a heuristic, with the ability to significantly reduce the action space, and hence training time. Detailed methodology along with its impact on convergence rate of the agent will be elaborated in the future research.}  %{ such that the number of selection possibilities reduces from ${\vert N^{gen} \vert}\choose{g}$ to $g \cdot {{\vert N^{gen}\vert/g} \choose{g}}$}{}. 
\end{itemize}

The heuristic and DRL training algorithm are described in following subsections.

\subsubsection{Heuristics} \label{heuristic} The heuristic rule {de-energizes  power system assets} proactively based on the %\change{minimum distance of approaching fire to the physical location of the equipment as defined in the reduced observation state $\hat{s}^f_k$}
{fire-distance metric} of the reduced observation space $\hat{S}^f$. 
If the {distance between the fire and} an asset $j \in \{N \cup T\}$ is less than or equal to a pre-defined threshold~$\beta$, then asset $j$ is de-energized, %it is deemed non-operational 
as defined in \eqref{eq:heurestics}. 
\begin{equation}
    z^{e}_{j,k} = 
    \begin{cases}
        0 & \text{if} \quad \hat{s}^{j}_{k} \leq \beta \\ 
        1 & \text{otherwise} 
    \end{cases}
    \label{eq:heurestics}
\end{equation}
Ideally, the value of $\beta$ depends upon the thermal characteristics of the equipment and the rate of temperature increase in the vicinity of the equipment's physical location. However, in this work, {since the} stochastic wildfire propagation model  does not explicitly capture temperature change{,} we use a constant value $\beta = 2$ {(}i.e., if the fire is 2 cells away from an asset $j$, then the corresponding external control input $z^{e}_{j,k}$ is set to 0{)}.

\subsubsection{DRL Training Algorithm}
\label{drl_algo}
To solve the problem described in Section~\ref{sec:Formulation}, we closely follow the Deep Deterministic Policy Gradient (DDPG)~\cite{lillicrap2015continuous} algorithm for the agent. The algorithm is a model-free, off-policy actor-critic algorithm using deep function approximators that can learn policies in high-dimensional, continuous action spaces.

The agent takes an action $a_k$ at every time step $k$ based on its trained policy $\mu$, $a_k = \mu(s_k|\theta^\mu)$, where $\theta^\mu$ is the weights of the agent (actor) for policy $\mu$. During training, the agent adds noise with the action it takes to explore the network. The actor updates its weights $\theta^\mu$ based on the critic value $Q(s, a|\theta^Q)$. The agent also uses target actor $\mu'$ and critic $Q'$ networks for stability. The target networks are initialized with the same random weights that $\theta^{Q'} \leftarrow \theta^Q$ and $\theta^{\mu'} \leftarrow \theta^\mu$.

The agent first calculates the expected return, $y_k = r_k + \gamma Q'(s_{k+1}, \mu'(s_{k+1}|\theta^{\mu'})|\theta^{Q'})$, based on the target networks, where $r_k$ is the reward value returned from the environment calculated based on \eqref{eq:total} and $\gamma$ is the discount factor for future rewards. Here, $y_k$ is called the moving target that the critic model tries to achieve. The agent updates the critic network $\theta^Q$ by minimizing the loss, $L=\frac{1}{N}\sum_k(y_k - Q(s_k, a_k | \theta^Q))^2$.

The actor network $\theta^\mu$ is updated based on the critic value. To calculate the actor loss, the agent first selects the action $a_k$ based on the actor policy $\mu$ and then applies the action to the critic network. The agent updates the actor policy {by} applying the sampled policy gradient, $\nabla_{\theta\mu}J \approx \frac{1}{N} \sum_k \nabla_aQ(s, a|\theta^Q)|_{s=s_k, a=\mu(s_k)}\nabla_{\theta\mu}\mu(s|\theta^\mu)|s_k $. 

Finally, the agent updates the target actor $\theta^{\mu'} = \tau\theta^\mu + (1-\tau)\theta^{\mu'}$ and target critic $\theta^{Q'} = \tau\theta^Q + (1-\tau)\theta^{Q'}$ networks, where $\tau \in (0, 1)$ is the target-network update constant.
To stabilize the learning process, constant $\tau$ is chosen to be a very small value, which prevents divergence due to fast updates. %a very small non-negative  constant, called, which is much less than one.\sm{this sentence does not make sense to me} This way the target network updates slowly which makes the network stable.

\section{Simulation Results} 
\label{simulation_result}

\subsection{Simulation Setup}
A standard IEEE 24-bus reliability test system (RTS), superimposed on a hypothetical geospatial terrain, divided into $350\times 350$ grid, has been considered here for analysis. Power system operational parameters can be obtained from \cite{4113721}. As discussed earlier, to reduce the size of the solution space, the outputs of these generators are aggregated. \change{Subsequently, there are ten controllable power generators.}{} Entire power system control horizon is divided into several time steps with a 15-minute interval. We consider the time step for the fire propagation is six times faster than power system operation. Here, every episode consists of 300 time steps (spanning the entire power system control horizon) or until we reach the non-convergence condition.  

% A standard IEEE 24-bus reliability test system (RTS), superimposed on a hypothetical geospatial terrain, divided into $350\times 350$ grid, has been considered here for analysis. Power system operational parameters can be obtained from \cite{4113721}. As discussed earlier, to reduce the size of the solution space, the outputs of these generators are aggregated. \change{Subsequently, there are ten controllable power generators.}{} Entire power system control horizon is divided into several time steps with 15-minute interval. We consider time step for the fire propagation is 6 times faster than power system operation. \change{So, the fire system state gets updated once after updating the power system six times.}{} \change{Given a deployed policy, e}{Here, e}very episode consists of 300 time steps \change{, corresponding to $\approx$ 3 days,}{(spanning the entire power system control horizon)} or until we reach the non-convergence condition.  

For simplicity, we also consider that the load demand within the power network is deterministic in nature with constant magnitude. As discussed, network-wide loads are comprised of critical and non-critical fractions. This fraction also remains constant throughout the network for simplicity. Our proposed controller tracks spatio-temporal wildfire propagation and provides set points for a partial set of generators. Successively, the power system operator would  solve the optimization problem to calculate and deploy the complete set of requisite control actions based on load forecasts. To determine the setpoints for the operator, we utilize the SCIP solver in \textit{General Algebraic Modeling System (GAMS)} due to its versatility. As discussed, non-convergence of power system operational problem is treated as a \textit{simulation-ending} condition with a severe penalty. To calculate the total penalty (see \eqref{eq:total}), constants $c_1$, $c_2$, $c_3$, and $c_4$ (corresponding to value of lost load, PIAW, ADIW, and non-convergence conditions) are 1, 1, 2, and 10, respectively.

We conduct our initial experiments considering a given wildfire origin, which remains the same during training and evaluation, % with the fire propagation depending on the transition probabilities, 
and we refer to this as a \textit{fixed source}. Successively, we expand our experiments by considering a random wildfire origin within the geographical area, which is chosen at random for each episode, and we refer to this as a \textit{random source}. We set a square boundary of $250 \times 250$ cells inside the $350 \times 350$ grid for the \textit{random source} of wildfire to ensure that the wildfire impacts the power system assets.

% We conduct our initial experiment \change{based on a fixed fire source}{considering a given wildfire origin with the fire propagation depending on the transition probabilities}, and we call this as a \textit{fixed-source}\change{, where the fire generation source remains fixed and propagates based on inherent dynamics}{}. Successively, we expand our experiment \change{for the random fire source}{considering the wild-fire originating randomly within the geographical terrain}, and we call this as a \textit{random-source}\change{, where the wild-fire can originate randomly within the geographical terrain. To ensure that the fire catches the power system components, we always start the fire somewhere in the middle}{}. We identify a square boundary of $100 \times 250$ inside the $350 \times 350$ grid \change{}{for \textit{random-source} of wildfire to ensure that wildfire actually impacts power system assets within the power system operational control horizon to expedite the training process}. \change{The fire can start anywhere inside the border.}{}

We use a deep neural network with two hidden layers, consisting of 450 and 300 neurons respectively, for both the actor and the critic network of the agent. %, % \change{to train the controller}{for training}, 
%where the first and second hidden layers consist of 450 and 300 neurons, respectively. 
In the actor network,
we use \emph{tanh} activation function in the hidden layers and \emph{sigmoid} activation  in the output layer.  
In the critic network, we use \emph{rectified linear} activation in the hidden layers and \emph{linear} activation in the output layer.  
The actor and critic learning rates are 0.001 and 0.002, respectively. 
We set 0.01 for the constant $\tau$ to update the target network and 0.99 as the discount factor $\gamma$ to calculate the expected return described in Section~\ref{drl_algo}.

\subsection{Control Approaches}
We consider three different control approaches in this experiment, which are defined as follows:

\subsubsection{Reactive control} 
In this approach, the system operator determines appropriate asset de-energization sequence, load shedding, and generator setpoints to be deployed based on its current observation of the interaction of wildfire with the power system (monitored transmission line status and emergency calls from substations) and predicted system-wide load demand. Here, the operator is blind to the wildfire propagation in the topographical space. External decision support is unavailable here.

% \change{We do not consider any external decision-making agent in this approach. As a result, the power system operator becomes blind to wildfire propagation.}{In this approach, the system operator determines appropriate asset deenergization sequence, load shedding and generator set points based its current observation of the interaction of wildfire with the power system (monitored transmission line status and emergency calls from the substation) and predicted system-wide load-demand. Here, the operator is blind to the wildfire propagation in the topographical space. External decision support is unavailable here.} \change{The operator can only react to monitored transmission line status and emergency calls from the substation, as and when wildfire interacts with them. Here, based on the current state of the power system, the operator can only provide remedial action in the next time step.}{}

\subsubsection{Myopic control} 
The external decision-making agent, based on its observation of wildfire propagation over the topographical space, provides the operator with proactive power system asset de-energization control input. The operator also observes the impact of wildfire on the power system, estimates system wide-load demand, and considers de-energization decision support from the controller to determine the control input for the next time step.

% \change{Compared to the reactive control, the operator gets help from the agent. Although, the agent is myopic in this case.}{} \change{The agent receives assistance only from the developed heuristic in decision-making.}{Here, the external decision making agent, based on its observation of wildfire propagation on the topographical space, provides the operator with proactive power system asset deenergization control input.} \change{Therefore, the operator is expected to able to save the system from wildfire directly interacting with transmission lines and substations. The operator still needs to calculate the generation and deployment of generation set-points for the next time step.}{The operator also observes the impact of wildfire on the power system, estimates system wide-load demand, and considers deenergization decision support from the controller, to determine the control input for the next time step.}

\subsubsection{Proactive control} 
In proactive control, the controller provides the operator with (i) power system asset de-energization control input based on the heuristic defined in Section \ref{heuristic}, and (ii) set points of the generators utilizing an RL-based controller. The RL-based controller can provide setpoints for a partial set or, entire generator fleet to the operator, and corresponding controllers are identified as full-control and partial-control, as described in Section \ref{reducing_actions}. The operator is also expected to observe the impact of wildfire on the power system, estimate system wide-load demand, and account for external control input, to calculate and deploy the requisite control actions.

\subsection{Training the Proactive Control Agent}
Training of DRL-based controllers can often be time-consuming, and it may take several days for an agent to learn a satisfactory policy. The pace of learning can also vary based on the random seed and selected hyper-parameter values. 

% \change{One of the time-consuming parts is to train the RL-based controller agent. It}{Training of RL-based controller can often be time consuming, and may take several days to} \change{obtain the}{learn the required} policy.\change{train the agent and get the test results.}{} The learning rate \change{of the agent may}{can also} vary based on the \change{random seed value}{selected hyper-parameters}. \change{Therefore, we ran the same five agents every time but using different seeds.}{}

% The agent explores \change{the network}{the controller solution space} by adding \change{}{random} noise with the actor-generated action. \change{So, we cannot surely tell how much the agent has learned based on penalties.}{Therefore, penalties along cannot indicate learning rate of the agent.} \change{On the other side, the agent's performance may decrease by overfitting.}{Additionally, overfitting may significantly impact agent's performance.} Therefore, we tested the agent for four episodes at every 20th episode to identify the best-trained agent. We also saved the trained agent at that position. Finally, we chose several trained agents based on the initial test results. Then, we tested it for 100 episodes. \sm{when you train, and when you test?} The agent is trained considering fixed- as well as the random-source.

During training, the agent explores the solution space by adding random noise to the actions chosen by the actor. 
Therefore, penalties received during  training cannot indicate the actual performance of the agent. Further, overfitting may %significantly impair 
diminish the agent's performance. In light of this, we evaluated the agent for 4 episodes (without exploration) after every 20 episodes of training to find the best version of the agent. 
After training has finished, %We also saved the trained agent at that position. Finally, 
we chose a set of trained agents based on these initial evaluations. Finally, we evaluated them thoroughly over 100 episodes.  
We trained agents in this manner %Note that agents were trained 
for both the fixed- and the random-source environments.

\subsection{Results}

\subsubsection{Fixed fire source}

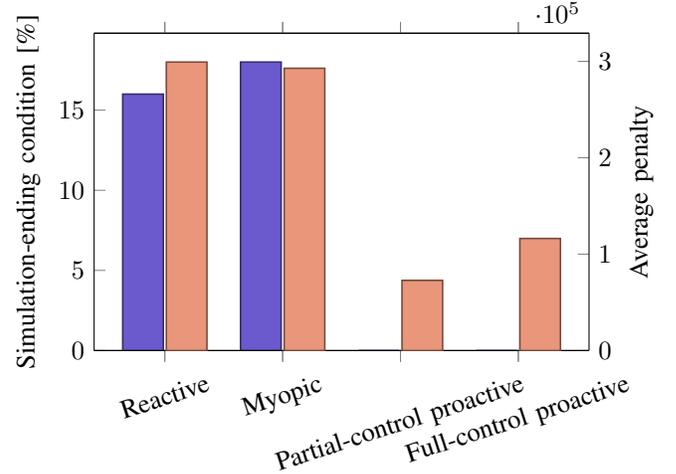
\begin{figure}[t]
\centering
\definecolor{color1}{RGB}{106,90,205}
\definecolor{color2}{RGB}{233,150,122}

\pgfplotsset{
    width=.45\textwidth,
    height=.32\textwidth
}

\begin{tikzpicture}
  \begin{axis}[
    ybar,
    area legend,
    symbolic x coords={Reactive, Myopic, Partial-control proactive, Full-control proactive},
    bar width=0.54cm, 
    bar shift=-0.29cm,
    enlarge x limits = 0.2,
    xtick=data,
    axis y line*=left,
    ymin=0,
    ylabel={Simulation-ending condition [\%]},
    x tick label style={white, rotate=18},
    ]
    \addplot[draw=color1!50!black, semithick, fill=color1] coordinates {
        (Reactive, 16) (Myopic, 18) (Partial-control proactive, 0) (Full-control proactive, 0)
      }; \label{clr1}
  \end{axis}
  \begin{axis}[
    ybar,
    area legend,
    symbolic x coords={Reactive, Myopic, Partial-control proactive, Full-control proactive},
    bar width=0.54cm, 
    bar shift=0.29cm,
    enlarge x limits=0.2,
    xtick=data,
    axis y line*=right,
    ymin=0,
    ylabel=Average penalty,
    x tick label style={rotate=18},
    ]
    \addplot[draw=color2!50!black, semithick, fill=color2] coordinates {
        (Reactive, 299239.74) (Myopic, 292816.67) (Partial-control proactive, 72799.22) (Full-control proactive, 116248.54)
      }; \label{clr2}
  \end{axis}
\end{tikzpicture}

\caption{Ratio of episodes that encountered a simulation-ending condition (\ref{clr1}) and average total penalty (\ref{clr2}) for various control approaches (reactive, myopic, partial-control proactive, %: fixed-source partial-control, 
and full-control proactive%: fixed-source full-control
) with fixed fire source. 
}
\label{fig:average_and_violations}
\end{figure}
As shown in Fig.~\ref{fig:average_and_violations}, the operator encounters no simulation-ending condition with the proactive control approach, while the reactive and myopic control approaches encounter simulation-ending states in 16\% and 18\% of episodes, respectively. The reason for encountering simulation-ending states is that the operator does not consider the entire time horizon when switching transmission assets {and providing set points to the generators} with these two approaches. 
This problem is addressed by the proactive control approach, as the controller is inherently trained to consider possible future states and penalties. % and obtain necessary actions for deployment. 
In Fig.~\ref{fig:average_and_violations}, we can also see that the average penalty is much lower for the proactive control agent (measured over 100 episodes).

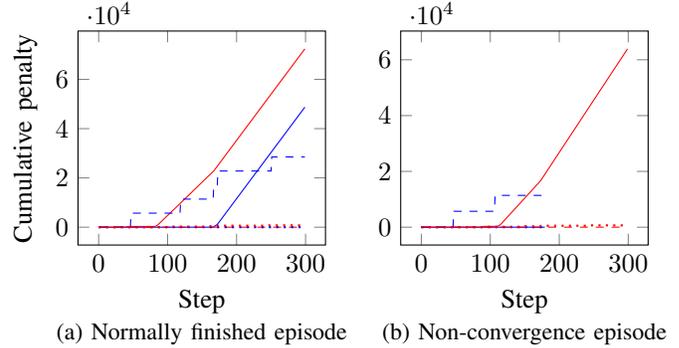
\begin{figure} [t]
\centering

\begin{tikzpicture}
\begin{groupplot}[
    group style={
    group name=total penalty,
    group size=2 by 1,
    xlabels at=edge bottom,
    ylabels at=edge left,
    horizontal sep=1cm,vertical sep=3cm,
    },
    legend columns=1, 
    legend style={at={(1.95, 1.7)},anchor=east, 
    /tikz/column 4/.style={
                column sep=15pt,
            },
    },
    width=0.55\linewidth,
    height=0.5\linewidth
    ]
    
    \nextgroupplot
    [
        xlabel style={align=center},
        xlabel={Step\\\small (a) Normally finished episode},
        ylabel={Cumulative penalty},
        mark repeat=25,
    ]
    
    % \node [text width=5em]  {\subcaption{\label{fig:penalty_component_example_episode}}};

    % \addlegendentry{load loss (reactive)}
    \addplot [sharp plot, blue] table[x=step, y=load_loss, col sep=comma] {test_results/episodic_analysis/general_case/reactive_cumulative.csv}; \label{r1}

    % \addlegendentry{load loss (proactive)}
    \addplot [sharp plot, red] table[x=step, y=load_loss, col sep=comma]
    {test_results/episodic_analysis/general_case/proactive_cumulative.csv}; \label{p1}
    
    % \addlegendentry{ADIW (reactive)}
    \addplot [sharp plot, blue, dashed] table[x=step, y=wildfire, col sep=comma] {test_results/episodic_analysis/general_case/reactive_cumulative.csv}; \label{r2}

    % \addlegendentry{ADIW (proactive)}
    \addplot [sharp plot, red, dashed] table[x=step, y=wildfire, col sep=comma]
    {test_results/episodic_analysis/general_case/proactive_cumulative.csv}; \label{p2}
    
    % \addlegendentry{PIAW (reactive)}
    \addplot [sharp plot, blue, dotted, thick] table[x=step, y=active_line_removal, col sep=comma] {test_results/episodic_analysis/general_case/reactive_cumulative.csv}; \label{r3}

    % \addlegendentry{PIAW (proactive)}
    \addplot [sharp plot, red, dotted, thick] table[x=step, y=active_line_removal, col sep=comma]
    {test_results/episodic_analysis/general_case/proactive_cumulative.csv}; \label{p3}

    % =============== non-convergence =============
    
    \nextgroupplot
    [
        xlabel style={align=center},
        xlabel={Step\\\small (b) Non-convergence episode},
        mark repeat=25,
    ]
    
    % \node [text width=5em]  {\subcaption{\label{fig:penalty_component_nc_example}}};
    
    % \addlegendentry{load loss (reactive)}
    \addplot [sharp plot, blue] table[x=step, y=load_loss, col sep=comma] {test_results/episodic_analysis/non_convergence/reactive_cumulative.csv};

    % \addlegendentry{load loss (proactive)}
    \addplot [sharp plot, red] table[x=step, y=load_loss, col sep=comma]
    {test_results/episodic_analysis/non_convergence/proactive_cumulative.csv};
    
    % \addlegendentry{wildfire (reactive)}
    \addplot [sharp plot, blue, dashed] table[x=step, y=wildfire, col sep=comma] {test_results/episodic_analysis/non_convergence/reactive_cumulative.csv};

    % \addlegendentry{wildfire (proactive)}
    \addplot [sharp plot, red, dashed] table[x=step, y=wildfire, col sep=comma]
    {test_results/episodic_analysis/non_convergence/proactive_cumulative.csv};
    
    % \addlegendentry{active line removal (reactive)}
    \addplot [sharp plot, blue, dotted, thick] table[x=step, y=active_line_removal, col sep=comma] {test_results/episodic_analysis/non_convergence/reactive_cumulative.csv};

    % \addlegendentry{active line removal (proactive)}
    \addplot [sharp plot, red, dotted, thick] table[x=step, y=active_line_removal, col sep=comma]
    {test_results/episodic_analysis/non_convergence/proactive_cumulative.csv};

\end{groupplot} 
\end{tikzpicture}
\caption{Cumulative penalty components for reactive (load loss (\ref{r1}), ADIW (\ref{r2}) and PIAW (\ref{r3})) and proactive (load loss (\ref{p1}), ADIW (\ref{p2}) and PIAW (\ref{p3})) control approaches over each step of two example episodes.}
\label{fig:penalty_component_example_episode}
\end{figure}

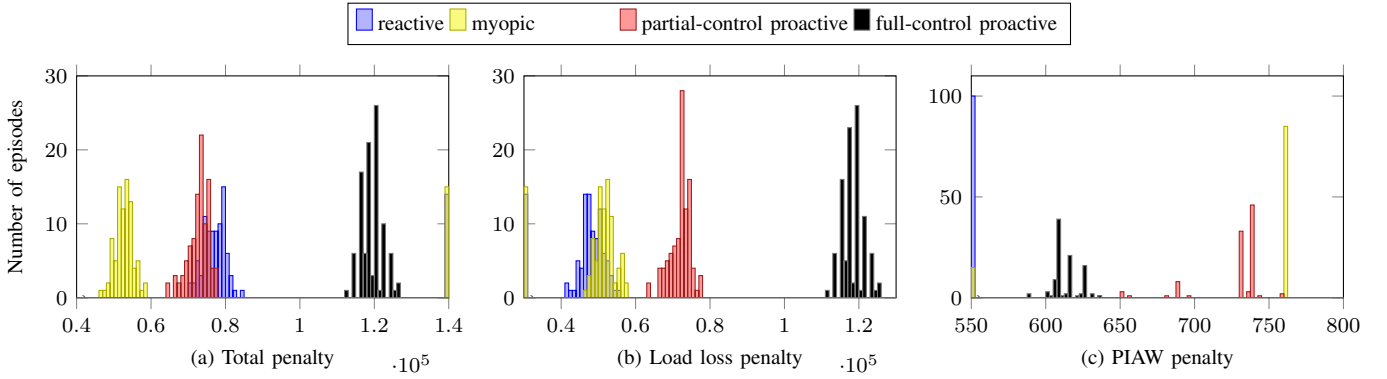
\begin{figure*}[!ht]
\centering

\definecolor{fixed-source-partial} {RGB}{255,50,50}
\definecolor{fixed-source-full}{RGB}{0,0,0}

\definecolor{myopic}{RGB}{245,245,0}

\definecolor{draw-line}{RGB}{230,230,230}

\begin{tikzpicture}
\begin{groupplot}[
    group style={
    group name=total penalty,
    group size=3 by 1,
    xlabels at=edge bottom,
    ylabels at=edge left,
    horizontal sep=1cm,vertical sep=3cm,
    },
    legend columns=4, 
    legend style={at={(1.7, 1.33)},anchor=north, 
    /tikz/column 4/.style={
                column sep=30pt,
            },
    },
    width=0.36\linewidth,
    height=0.25\linewidth
    ]
    
    % ============ Total Penalty ============
\nextgroupplot
    [
    font=\footnotesize,
    ybar,
    bar width=20pt,
    xmin=40000,
    xmax=140000,
    ymin=0,
    ymax=30,
    xlabel={(a) Total penalty},
    ylabel={Number of episodes},
    ]
    
    \node [text width=5em]  {\subcaption{\label{fig:baseline_and_rl_agents_a}}};
    \addlegendentry{reactive}
    
    \addplot +[
        hist={
            bins=100,
        }   
    ] table [x=episode_number, y=total_penalty, col sep=comma]     {test_results/base_case/reactive.csv};

    \addlegendentry{myopic}
    \addplot +[
        hist={
            bins=100,
        },
        myopic,
        draw=myopic!70!black,
        fill opacity=0.5
    ] table [x=episode_number, y=total_penalty, col sep=comma]     {test_results/base_case/myopic.csv};

    \addlegendentry{partial-control proactive}
    \addplot +[
        hist={
            bins=100,
        },        
        fill opacity=0.5,
        fixed-source-partial,
        draw=fixed-source-partial!60!black
    ] table [x=episode_number, y=total_penalty, col sep=comma]     {test_results/partial_control_over_power_generation/fixed_fire_source.csv};

    \addlegendentry{full-control proactive}
    \addplot +[
        hist={
            bins=100,
        },
        fixed-source-full,
        draw=fixed-source-full!50!draw-line
    ] table [x=episode_number, y=total_penalty, col sep=comma]     {test_results/full_control_over_power_generation/fixed_fire_source.csv};

    % ============= load loss =================
\nextgroupplot
    [
    font=\footnotesize,
    ybar,
    bar width=20pt,
    xmin=30000,
    xmax=130000,
    ymin=0,
    ymax=30,
    xlabel={(b) Load loss penalty},
    ]
    
    \node [text width=5em]  {\subcaption{\label{fig:baseline_and_rl_agents_b}}};

    % \addlegendentry{baseline}
    \addplot +[
        hist={
            bins=100,
        }   
    ] table [x=episode_number, y=load_loss, col sep=comma]     {test_results/base_case/reactive.csv};

    % \addlegendentry{myopic}
    \addplot +[
        hist={
            bins=100,
        }, 
        myopic,
        draw=myopic!70!black,
        fill opacity=0.5   
    ] table [x=episode_number, y=load_loss, col sep=comma]     {test_results/base_case/myopic.csv};

    % \addlegendentry{RL partial-control}
    \addplot +[
        hist={
            bins=100,
        },        
        fill opacity=0.5,
        fixed-source-partial,
        draw=fixed-source-partial!60!black
    ] table [x=episode_number, y=load_loss, col sep=comma]     {test_results/partial_control_over_power_generation/fixed_fire_source.csv};

    % \addlegendentry{RL full-control}
    \addplot +[
        hist={
            bins=100,
        },
        fixed-source-full,
        draw=fixed-source-full!50!draw-line
    ] table [x=episode_number, y=load_loss, col sep=comma]     {test_results/full_control_over_power_generation/fixed_fire_source.csv};

    % ============== active line removal ===========
    
\nextgroupplot
    [
    font=\footnotesize,
    ybar,
    bar width=20pt,
    xmin=550,
    xmax=800,
    ymin=0,
    xlabel={(c) PIAW penalty},
    ]

    \node [text width=5em]  {\subcaption{\label{fig:baseline_and_rl_agents_c}}};

    % \addlegendentry{baseline}
    \addplot +[
        hist={
            bins=100,
        }   
    ] table [x=episode_number, y=active_line_removal, col sep=comma]     {test_results/base_case/reactive.csv};

    % \addlegendentry{myopic}
    \addplot +[
        hist={
            bins=100,
        },        
        myopic,
        draw=myopic!70!black,
        fill opacity=0.5   
    ] table [x=episode_number, y=active_line_removal, col sep=comma]     {test_results/base_case/myopic.csv};

    % \addlegendentry{RL partial-control}
    \addplot +[
        hist={
            bins=100,
        },        
        fill opacity=0.5,
        fixed-source-partial,
        draw=fixed-source-partial!60!black
    ] table [x=episode_number, y=active_line_removal, col sep=comma]     {test_results/partial_control_over_power_generation/fixed_fire_source.csv};

    % \addlegendentry{RL full-control}
    \addplot +[
        hist={
            bins=100,
        },
        fixed-source-full,
        draw=fixed-source-full!50!draw-line
    ] table [x=episode_number, y=active_line_removal, col sep=comma]     {test_results/full_control_over_power_generation/fixed_fire_source.csv};
\end{groupplot}

\end{tikzpicture}
\caption{Distribution of total penalty over 100 episodes for different types of controls (reactive, myopic, partial-control proactive, %: fixed-source partial-control, 
and full-control proactive%: fixed-source full-control
) with a fixed fire source.
% \todo[inline]{would be good to mention how many episodes are plotted here}
% \textcolor{red}{same as before, instead of 1 and 2, refer to them as Partial Proactive and Full Proactive (or something similarly meaningful)}
}
\label{fig:baseline_and_rl_agents}
\end{figure*}

\begin{figure*}[!ht]
\centering

\definecolor{fixed-source-partial} {RGB}{255,50,50}
\definecolor{fixed-source-full}{RGB}{0,0,0}

\definecolor{random-source-full}{RGB}{0,255,255}
\definecolor{random-source-partial}{RGB}{0,235,0}

\definecolor{draw-line}{RGB}{230,230,230}

\begin{tikzpicture}
\begin{groupplot}[
    group style={
    group name=total penalty,
    group size=3 by 1,
    xlabels at=edge bottom,
    ylabels at=edge left,
    horizontal sep=1cm,vertical sep=3cm,
    },
    legend columns=4, 
    legend style={at={(1.7, 1.33)},anchor=north, 
    /tikz/column 4/.style={
                column sep=15pt,
            },
    },
    width=0.36\linewidth,
    height=0.25\linewidth
    ]
    
    % ============ Total Penalty ============
\nextgroupplot
    [
    font=\footnotesize,
    ybar,
    bar width=20pt,
    xmin=50000,
    xmax=220000,
    ymin=0,
    ymax=40,
    xlabel={(a) Total penalty},
    ylabel={Number of episodes},
    ]

    \node [text width=5em]  {\subcaption{\label{fig:group_only_rl_agents_a}}};

    \addlegendentry{fixed-source partial-control}
    \addplot +[
        hist={
            bins=100,
        }, 
        fixed-source-partial,
        draw=fixed-source-partial!60!black
    ] table [x=episode_number, y=total_penalty, col sep=comma]     {test_results/partial_control_over_power_generation/fixed_fire_source.csv};

    \addlegendentry{fixed-source full-control}
    \addplot +[
        hist={ 
            bins=100,
        }, 
        fixed-source-full,
        draw=fixed-source-full!50!draw-line
    ] table [x=episode_number, y=total_penalty, col sep=comma]     {test_results/full_control_over_power_generation/fixed_fire_source.csv};

    \addlegendentry{random-source full-control}
    \addplot +[
        hist={
            bins=100,
        }, 
        random-source-full,
        draw=random-source-full!50!black
    ] table [x=episode_number, y=total_penalty, col sep=comma]     {test_results/full_control_over_power_generation/random_fire_source.csv};

    \addlegendentry{random-source partial-control}
    \addplot +[
        hist={
            bins=100,
        }, 
        random-source-partial,
        fill opacity=0.5,
        draw=random-source-partial!50!black
    ] table [x=episode_number, y=total_penalty, col sep=comma]     {test_results/partial_control_over_power_generation/random_fire_source.csv};

    % ============= load loss =================
\nextgroupplot
    [
    font=\footnotesize,
    ybar,
    bar width=20pt,
    xmin=50000,
    xmax=220000,
    ymin=0,
    ymax=40,
    xlabel={(b) Load loss penalty},
    ]
    
    \node [text width=5em]  {\subcaption{\label{fig:group_only_rl_agents_b}}};

    % \addlegendentry{fixed-source partial-control}
    \addplot +[
        % fixed-source-partial,
        hist={
            bins=100,
        },
        fixed-source-partial,
        draw=fixed-source-partial!60!black
    ] table [x=episode_number, y=load_loss, col sep=comma]     {test_results/partial_control_over_power_generation/fixed_fire_source.csv};

    % \addlegendentry{fixed-source full-control}
    \addplot +[
        % fixed-source-full,
        hist={
            bins=100,
        }, 
        fixed-source-full,
        draw=fixed-source-full!50!draw-line
    ] table [x=episode_number, y=load_loss, col sep=comma]     {test_results/full_control_over_power_generation/fixed_fire_source.csv};

    % \addlegendentry{random-source full-control}
    \addplot +[
        % random-source-full,
        hist={
            bins=100,
        },
        random-source-full,
        draw=random-source-full!50!black
    ] table [x=episode_number, y=load_loss, col sep=comma]     {test_results/full_control_over_power_generation/random_fire_source.csv};

    % \addlegendentry{random-source partial-control}
    \addplot +[
        % random-source-partial,
        hist={
            bins=100,
        },
        random-source-partial,
        fill opacity=0.5,
        draw=random-source-partial!50!black
    ] table [x=episode_number, y=load_loss, col sep=comma]     {test_results/partial_control_over_power_generation/random_fire_source.csv};
    
    % ============== active line removal ===========
    
 \nextgroupplot
    [
    font=\footnotesize,
    ybar,
    bar width=20pt,
    xmin=-50,
    xmax=1350,
    ymin=0,
    xlabel={(c) PIAW penalty},
    ]
    
    \node [text width=5em]  {\subcaption{\label{fig:group_only_rl_agents_c}}};

    % \addlegendentry{fixed-source partial-control}
    \addplot +[
        hist={
            bins=100,
        },
        fixed-source-partial,
        draw=fixed-source-partial!60!black
    ] table [x=episode_number, y=active_line_removal, col sep=comma]     {test_results/partial_control_over_power_generation/fixed_fire_source.csv};

    % \addlegendentry{fixed-source full-control}
    \addplot +[
        hist={
            bins=100,
        },
        fixed-source-full,
        draw=fixed-source-full!50!draw-line
    ] table [x=episode_number, y=active_line_removal, col sep=comma]     {test_results/full_control_over_power_generation/fixed_fire_source.csv};

    % \addlegendentry{random-source full-control}
    \addplot +[ 
        hist={
            bins=100,
        },
        random-source-full,
        draw=random-source-full!50!black
    ] table [x=episode_number, y=active_line_removal, col sep=comma]     {test_results/full_control_over_power_generation/random_fire_source.csv};

    % \addlegendentry{random-source partial-control}
    \addplot +[
        hist={
            bins=100,
        },
        random-source-partial,
        fill opacity=0.5,
        draw=random-source-partial!50!black
    ] table [x=episode_number, y=active_line_removal, col sep=comma]     {test_results/partial_control_over_power_generation/random_fire_source.csv};

\end{groupplot} 

% \node[text width=6cm,align=center,anchor=north] at ([yshift=-5mm]total penalty c1r1.south) {\captionof{subfigure}{(A) total penalty \label{subplot:one}}};

% \node[text width=6cm,align=center,anchor=north] at ([yshift=-5mm]total penalty c2r1.south) {\captionof{subfigure}{(A) total penalty \label{subplot:two}}};

\end{tikzpicture}
\caption{Distribution of penalty over 100 episodes for various types of proactive control with fixed and random fire sources.
% \todo[inline]{would be good to mention how many episodes are plotted here}
}
\label{fig:group_only_rl_agents}
\end{figure*}
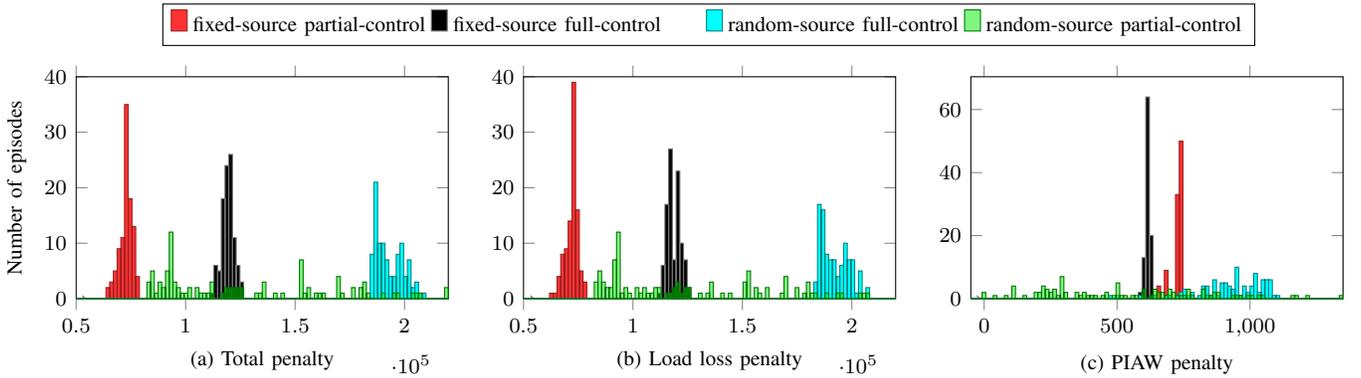

Figs. \ref{fig:penalty_component_example_episode}a and~\ref{fig:penalty_component_example_episode}b show temporal variation of various penalty component (without \textit{non-convergence}) for given wildfire origin and propagation paths. Both Figs.~\ref{fig:penalty_component_example_episode}a and \ref{fig:penalty_component_example_episode}b shows that although the load loss penalty in the proactive control approach is higher compared to the reactive control approach, this is compensated by other components of the reward function, indicating reduced flow through the `to be' de-energized lines or alleviation of non-convergence. It is important to note that load loss penalty in the proactive control approach need not always be higher compared to the reactive control approach. Fig.~\ref{fig:penalty_component_example_episode}(b) shows that the ability to estimate the future and accounting for it in the decision making, even with the myopic operator, ensures that the system operation does not end prematurely due to non-convergence.

In Figure \ref{fig:baseline_and_rl_agents_a}, we see the distribution of episodic \textit{total penalty} for different types of control approaches. Since the total load within the system remains constant, the non-convergence penalty, if encountered, will also be constant. Therefore, we did not include the distribution of the non-convergence penalty, and Fig. \ref{fig:average_and_violations} can be referred to in this regard. It can be seen from the figures that the median of the total penalty for the proactive partial-control agent falls in between the reactive control approach and the myopic control agent if non-convergence penalties are ignored. The ability to avoid non-convergence with a proactive control agent indicates the superiority of this approach. It can also be seen that partial control works better than controlling the entire generating fleet. %\change{This can be due to the power system operator maintains the power flow.}{}

% In Figure \ref{fig:baseline_and_rl_agents_a}, we see the distribution of episodic \textit{total penalty} \change{for the first 100 episodes}{} for different type\change{}{s} of control approaches. \change{Note that the figure does not}{Since, the total load within the system remains constant, non-convergence penalty, if encountered, will also be constant. Therefore, we did not} include the distribution of non-convergence penalty \change{.}{, and Fig. \ref{fig:average_and_violations} can be referred to in this regard}. \change{The proactive fixed-source full-control agent's median total penalty is around 120K and 72K for the proactive fixed-source partial-control agent. The median of the total penalty for the reactive control approaches is around 80K, where the myopic control agent's penalty is around 53K.}{} \change{So,}{It can be seen from the figures that} the median of the total penalty for the proactive \change{fixed-source}{} partial-control agent falls in between the reactive control approach and the myopic control agent if \change{we do not include the}{} non-convergence penalties \change{}{are ignored. Ability to avoid non-convergence with proactive control agent indicates superiority of this approach}. \change{We can also notice that it}{It can also be seen that partial control works better} \change{if the proactive control agent controls the power generation partially}{than controlling entire generating fleet. This can be due to overfitting by the agent.} 

Figs. \ref{fig:baseline_and_rl_agents_b} and  \ref{fig:baseline_and_rl_agents_c} show the load loss and the PIAW penalty, respectively. From the earlier discussion, it is evident that the proactive control agent tries to find equilibrium by reducing the power flow {through the assets, minimizing ADIW and PIAW and avoiding} non-convergence, {while also} minimizing load loss penalty. Note that the reactive control approach does not have a PIAW penalty as it does not remove lines proactively. Instead, it incurs an ADIW penalty for not taking any action.

% Fig\change{ure}{s.} \ref{fig:baseline_and_rl_agents_b} \change{shows the \textit{load loss} penalty and Figure~}{and } \ref{fig:baseline_and_rl_agents_c} show\change{s}{the \textit{load loss} and} the \textit{PIAW} penalty \change{}{respectively}. \change{As mentioned in Section~\ref{sec:Formulation}, the PIAW penalty is proportional to the amount of power that flows through the line. The amount of power that flows through the line depends upon the fire source and the propagation probability. For the fixed fire source, it is always the same. From Figure~\ref{fig:baseline_and_rl_agents_c}, we understand that the proactive control agent avoids the non-convergence penalty by reducing the amount of power that flows through the line. On the other hand, it increases the load loss penalty. We can see that the load loss penalty is maximum when the PIAW penalty is minimum.}{} \change{}{From the earlier discussion it is imminent that} the proactive control agent tries to find equilibrium by reducing the power flow to avoid the non-convergence but keeping the power flow high enough for a minimum load loss penalty. Note that the reactive control approach does not have an PIAW penalty as it does not remove lines proactively. Instead, it gets a \textit{ADIW} penalty for not taking any action. 

\subsubsection{Proactive control agents based on random fire sources}
We also trained and evaluated agents {considering} random fire sources, as shown in Fig. \ref{fig:group_only_rl_agents_a}. {Here,} both  proactive control agents can again learn to avoid simulation-ending conditions. Although the random-source partial-control agent seems better than the random-source full-control agent, the agent encountered simulation-ending conditions in 2\% of the cases. 

Note that in some of the figures, we see distribution columns at the edge of the figures. These columns represent all the values that are outside of the plotted range. % We kept those values out of range (intentionally) in the figures. 
In Figure \ref{fig:baseline_and_rl_agents_a}, we see a yellow column (myopic control agent) at the right edge. Those are the total penalties that include a non-convergence penalty for reaching simulation-ending conditions. Similar phenomenon can also be observed in Fig. \ref{fig:group_only_rl_agents_a}.

\section{Conclusions} 
\label{conclusions}
% {\color{red}[0.25 pages] Summarize key contributions and lessons learned, allude to future work.}

In this work, we  developed a deep reinforcement learning (DRL) based proactive control to supplement decision support for operators given a wildfire event. Testbed has been developed integrating a wildfire-propagation model with a power-system operation model to train and validate a controller that can supplement traditional computationally-intensive, forecast-driven power-system operations during a wildfire. 
We formulated the control problem as a Markov Decision Process (MDP). To address the challenges posed by the large observation space, we introduced a compact representation of fire-state observations. To tackle the challenges of the large action space, we proposed a hybrid agent, consisting of a heuristic and a deep neural network, which facilitate the training of the controller.
In particular, we utilized a model-free, off-policy, actor-critic algorithm using deep function approximators as a part of the Deep Deterministic
Policy Gradient (DDPG) algorithm to train a deep RL-based policy. We also aggregated generator outputs to reduce the size of the action space.

Numerical results indicate that the DRL-based proactive control agent can avoid thermal overloading of the transmission network when one or more transmission assets are outaged due to safety concerns with encroaching wildfire, while reactive and myopic approaches struggle to do so.
Since our agent needs to be trained before the wildfire event, a crucial question is whether it can perform well if the origin of the fire is unknown at training time: our results demonstrate that while it is easier to handle fixed-source fires, our agent performs well even with random fire sources.
%While the controller performs well with the fixed fire origin, its performance gets worsened when wildfire originates randomly within the topographical space. 
We also observed that our proactive control agent performs significantly better when it provides setpoints for only a subset of generators and lets the operator derive setpoints for the remaining ones (compared to providing setpoints to all generators). %instead of providing setpoints   compared to a proactive controller providing setpoints for all the generators, a controller providig setpoints for only a subset of generators  performs significantly better.  the performance significantly improves when it provides the control signal to a partial set of generators, and the operator derives set points for the remaining set. 
%Finally, we also described a methodology for real-world deployment\Aron{where was this?}\sm{It was shown in Fig. 1, and has been explained whereever needed}. 

%\change{In this work, we aggregated the outputs of  generators when there are 
%To reduce the size of \Aron{action space?}solution space, we aggregated the output of the generators when the IEEE 24-bus RTS contains 
%multiple generators at a given bus to reduce the size of the action space. }{}
\change{With respect to directions for future work, we conjecture that controlling the output of every generator individually---some of which we aggregated in this work to reduce the action space---might help us to control the flow through the outgoing assets in a better way and possibly reduce load shedding. AC-OPF could also be used in deriving the setpoints.}{} 
In this work, we  reduced the size of the observation % {and action} % REST OF THIS PARAGRAPH SEEMS TO PERTAIN ONLY TO THE OBSERVATION SPACE, SO LET'S NOT BRING UP THE ACTION SPACE HERE
%\Salah{I think we have described combining heuristic with DRL as a contribution. And also, using heuristic is not the reason of sub-optimal.}\Salah{Also, converting to fire distance from whole grid information (reducing observation space) is a contribution which helps to get faster and better result}
space {by utilizing heuristics} for the sake of computational tractability, which enables the application of our approach to large geographical areas. However, 
%Similarly, while the reduction of the size of the observation space is suitably reduced, the derived controller based on such reduced space may not be optimal. 
actions chosen based on such a reduced space might be sub-optimal.
In future work, we will explore approaches for handling large observation spaces, such as (graph) convolutional neural networks. % (CNNs) and graph CNNs.
\ifCLASSOPTIONcaptionsoff
  \newpage
\fi

% references section

% can use a bibliography generated by BibTeX as a .bbl file
% BibTeX documentation can be easily obtained at:
% http://mirror.ctan.org/biblio/bibtex/contrib/doc/
% The IEEEtran BibTeX style support page is at:
% http://www.michaelshell.org/tex/ieeetran/bibtex/

\bibliographystyle{IEEEtran}
% argument is your BibTeX string definitions and bibliography database(s)
\bibliography{7_references.bib}

\end{document}